\magnification=\magstephalf
\baselineskip=16pt
\parskip=8pt
\rightskip=0.5truecm

\def\a{\alpha}
\def\b{\beta}
\def\d{\delta}
\def\e{\epsilon}

\def\g{\gamma}
\def\l{\lambda}
\def\k{\kappa}
\def\r{\rho}
\def\s{\sigma}
\def\t{\tau}

\def\o{\omega}
\def\D{\Delta}
\def\L{\Lambda}
\def\G{\Gamma}
\def\O{\Omega}
\def\S{\Sigma}

\def\del #1{\frac{\partial^{#1}}{\partial\l^{#1}}}

\def\Const{Const\,}
\def\const{const\,}

\def\1{1}

\def\E{I\kern-.25em{E}}
\def\N{I\kern-.22em{N}}
\def\M{I\kern-.22em{M}}
\def\R{I\kern-.22em{R}}
\def\Q{I\kern-.22em{Q}}
\def\Z{{Z\kern-.5em{Z}}}

\def\C{C\kern-.75em{C}}
\def\P{I\kern-.25em{P}}

\def\del{\partial}


\def\EE{{\cal E}}

\def\HH{{\cal H}}

\def\NN{{\cal N}}

\def\RR{{\cal R}}

\def\chap #1{\line{\ch #1\hfill}}

\def\one{\hbox{J}\kern-.2em\hbox{I}}
\def\un #1{\underline{#1}}
\def\ov #1{\overline{#1}}
\def\ba{{\backslash}}
\def\sb{{\subset}}
\def\sp{{\supset}}

\def\em{{\emptyset}}


\newcount\foot
\foot=1
\def\note#1{\footnote{${}^{\number\foot}$}{\ftn #1}\advance\foot by 1}
\def\tag #1{\eqno{\hbox{\rm(#1)}}}
\def\frac#1#2{{#1\over #2}}
\def\text#1{\quad{\hbox{#1}}\quad}

\def\proposition #1{\noindent{\thbf Proposition #1: }}
\def\datei #1{\headline{\rm \the\day.\the\month.\the\year{}\hfill{#1.tex}}}

\def\theo #1{\noindent{\thbf Theorem #1: }}
\def\lemma #1{\noindent{\thbf Lemma #1: }}

\def\proof{{\noindent\pr Proof: }}

\def\endproof{$\diamondsuit$}
\def\remark{{\bf Remark: }}

\def\endproof{$\diamondsuit$}

\font\pr=cmbxsl10 scaled\magstephalf
\font\thbf=cmbxsl10 scaled\magstephalf

\long\def\fussnote#1#2{{\baselineskip=10pt
\setbox\strutbox=\hbox{\vrule height 7pt depth 2pt width 0pt}
\sevenrm
\footnote{#1}{#2}
}}


\font\ch=cmbx12

\font\ftn=cmr8

\font\it=cmti10
\font\bf=cmbx10
\font\srm=cmr5


\overfullrule=0pt
\font\tit=cmbx12
\font\aut=cmbx12
\font\aff=cmsl12
{$  $}
\vskip2truecm
\centerline{\tit 
STABILITY FOR A CONTINUOUS SOS-INTERFACE-MODEL
IN A 
}
\vskip.2truecm
\centerline{\tit RANDOMLY PERTURBED
PERIODIC POTENTIAL\footnote{${}^*$}{\ftn Work
supported by the DFG
Schwerpunkt `Stochastische Systeme hoher Komplexit\"at'
}}
\vskip2truecm
\vskip.5truecm
\centerline{\aut  Christof K\"ulske\footnote{${}^{1}$}{\ftn
e-mail: kuelske@wias-berlin.de}
}


\vskip.1truecm
\centerline{\aff WIAS}
\centerline{\aff Mohrenstrasse 39}
\centerline{\aff D-10117 Berlin, Germany}
\vskip1truecm\rm

\noindent {\bf Abstract:}
We consider the Gibbs-measures of continuous-valued height 
configurations on the $d$-dimensional integer lattice 
in the presence a weakly disordered potential.  
The potential is composed of
Gaussians having random location and random depth; it
becomes periodic under shift of the interface perpendicular to 
the base-plane for zero disorder. 
We prove that there exist localized interfaces with probability 
one in dimensions $d\geq 3+1$, in a `low-temperature' regime. 
The proof extends the method of continuous-to-discrete single-site 
coarse graining that was previously applied by the author 
for a double-well
potential to the case of a non-compact image space.
This allows to utilize parts of the renormalization group analysis 
developed for the treatment of a contour representation
of a related integer-valued SOS-model in [BoK1].
We show that, for a.e. fixed realization 
of the disorder, the infinite volume Gibbs measures 
then have a representation as superpositions of 
massive Gaussian fields with centerings that 
are distributed according to the infinite volume Gibbs measures
of the disordered integer-valued SOS-model with exponentially decaying
interactions. 



\vfill
     ${}$
\eject



\magnification=\magstephalf
\baselineskip=16pt
\parskip=8pt
\rightskip=0.5truecm

\def\a{\alpha}
\def\b{\beta}
\def\d{\delta}
\def\e{\epsilon}

\def\g{\gamma}
\def\l{\lambda}
\def\k{\kappa}
\def\r{\rho}
\def\s{\sigma}
\def\t{\tau}

\def\o{\omega}
\def\D{\Delta}
\def\L{\Lambda}
\def\G{\Gamma}
\def\O{\Omega}
\def\S{\Sigma}

\def\del #1{\frac{\partial^{#1}}{\partial\l^{#1}}}

\def\Const{Const\,}
\def\const{const\,}

\def\1{1}

\def\E{I\kern-.25em{E}}
\def\N{I\kern-.22em{N}}
\def\M{I\kern-.22em{M}}
\def\R{I\kern-.22em{R}}
\def\Z{{Z\kern-.5em{Z}}}

\def\C{C\kern-.75em{C}}
\def\P{I\kern-.25em{P}}

\def\del{\partial}


\def\EE{{\cal E}}

\def\HH{{\cal H}}

\def\NN{{\cal N}}

\def\RR{{\cal R}}

\def\chap #1{\line{\ch #1\hfill}}

\def\one{\hbox{J}\kern-.2em\hbox{I}}
\def\un #1{\underline{#1}}
\def\ov #1{\overline{#1}}
\def\ba{{\backslash}}
\def\sb{{\subset}}
\def\sp{{\supset}}

\def\em{{\emptyset}}


\newcount\foot
\foot=1
\def\note#1{\footnote{${}^{\number\foot}$}{\ftn #1}\advance\foot by 1}
\def\tag #1{\eqno{\hbox{\rm(#1)}}}
\def\frac#1#2{{#1\over #2}}
\def\text#1{\quad{\hbox{#1}}\quad}

\def\proposition #1{\noindent{\thbf Proposition #1: }}
\def\datei #1{\headline{\rm \the\day.\the\month.\the\year{}\hfill{#1.tex}}}

\def\theo #1{\noindent{\thbf Theorem #1: }}
\def\lemma #1{\noindent{\thbf Lemma #1: }}

\def\proof{{\noindent\pr Proof: }}

\def\endproof{$\diamondsuit$}
\def\remark{{\bf Remark: }}

\def\endproof{$\diamondsuit$}

\font\pr=cmbxsl10 scaled\magstephalf
\font\thbf=cmbxsl10 scaled\magstephalf

\long\def\fussnote#1#2{{\baselineskip=10pt
\setbox\strutbox=\hbox{\vrule height 7pt depth 2pt width 0pt}
\sevenrm
\footnote{#1}{#2}
}}


\font\ch=cmbx12

\font\ftn=cmr8

\font\it=cmti10
\font\bf=cmbx10
\font\srm=cmr5



\bigskip\bigskip

\chap{I. Introduction}

The study of interface models from statistical mechanics,
continuous as well as discrete ones, 
with respect to their localization vs. fluctuation properties,
is an interesting topic in probability theory. 
In this paper we study the problem of continuous 
SOS-interfaces in random potentials that are random 
perturbations of periodic ones and prove 
stability of the interface in dimensions $d\geq 3+1$ 
(as suggested by the heuristic Imry-Ma argument, known
for long to theoretical physicists).

A related stability result has been proved before 
for the simpler discrete version of such a model 
with nearest neighbor interactions in [BoK1]. 
The proof uses a renormalization group (or spatial coarse-graining)
procedure that was based on the technique of 
Bricmont and Kupiainen that was developed for the 
Random Field Ising Model [BK]. 
The issue of this note is thus to clarify what to do with  
additional (possibly destabilizing) 
fluctuations of the continuous degrees of freedom.

An analogous problem 
was investigated in a recent paper by the author [K4] 
in the simpler case of a random double-well potential, where
ferromagnetic ordering was shown in $d\geq 3$
(under suitable `low temperature' 
and `weak anharmonicity' assumptions on the potential).
The key point here is to construct a suitable stochastic
mapping from continuous to discrete configurations and
study the image measures under this mapping. 
In the double-well case this mapping is just a smoothed
sign-field indicating what minimum the 
continuous spin is close to. 
The image measure could then be shown to be an Ising-measure 
for a suitable absolutely summable
Hamiltonian. It can be controlled by the known renormalization group 
method of [BK]. 
This is clearly in favor of running a suitably devised 
renormalization  group transformation
on the (in this context unpleasantly rich) 
space of continuous configurations. 
 
The purpose of the present paper is to study the difficulties
of infinitely many minima in the potential.
The stochastic mapping we will apply to the 
continuous spins will now be a mapping to integer-
valued spin configurations. 
As opposed to [K4] the mapping will now also 
depend on the realization of the disorder. 

To explain the method in the simplest non-trivial context,
we have decided to choose a specific potential that 
is the log of sums of Gaussians. The treatment of this 
potential provides the basic building block of the analysis 
also for more general potentials
in that it explains the occurrence of the phase transition
and the structure of the contour models that will arise. 
It corresponds to having vanishing `anharmonic corrections';
how those anharmonicities (that are present for more 
general potentials)
can be treated by additional expansions is explained in detail 
for the double-well case in [K4], so that combining those
methods with the ones from the present paper should
yield stability for a larger class of continuous 
interface models.  

This restriction also allows us to obtain 
particularly nice `factorization-formulas' 
for the continuous-spin Gibbs-measures in finite and 
infinite volume. They have some probabilistic appeal
and clarify the structure of the coarse graining
transformation we use.
In particular we can describe 
the infinite volume Gibbs measures 
in terms of the `explicit' building blocks
of random discrete height measures
and well-understood (random) massive Gaussian fields (see [1.7]). 

Here is the model.
We are aiming to investigate the Gibbs measures 
on the state space $\R^{\Z^d}$
of the continuous spin model given by the 
Hamiltonians in finite volumes $\L\sb \Z^d$
$$
\eqalign{
&E^{\tilde m_{\del \L},\o_{\L}}_{\L}
\left(m_{\L}\right)
=\frac{q}{2}\sum_{{\{x,y\}\sb \L}\atop {d(x,y)=1}}\left(
m_x-m_y
\right)^2+
\frac{q}{2}\sum_{{x\in \L; y\in \del \L}
\atop{d(x,y)=1}}\left(
m_x-\tilde m_y
\right)^2 + \sum_{x\in \L}V_x(m_x)
}
\tag{1.1}
$$
for a configuration $m_{\L}\in \R^{\L}$
with boundary condition $\tilde m_{\del \L}$.
We write $\del \L=\{x\in \L^c;\exists y\in \L:  d(x,y)=1\}$
for the outer boundary of a set $\L$ where 
$d(x,y)=\Vert x-y\Vert_1$ is the $1$-norm on $\R^d$.
$q\geq 0$ will be small. 

The random potential we consider is given by 
$$
\eqalign{
&V_x(m_x)
=-\log\left[\sum_{l\in \Z}e^{
-\frac{1}{2}
\left(m_x-m^*_x(l)  \right)^2+\eta_x(l)}\right]\text{where}\cr
&m^*_x(l)=m^* \left(l+d_x(l)\right)\cr
}
\tag{1.2}
$$
The disorder is modelled by the random variables
$\left(\eta_x(h)\right)_{x\in \Z^d;h\in \Z}$ and
$\left(d_x(h)\right)_{x\in \Z^d;h\in \Z}$,
describing the random depths of the Gaussians
and the random deviations
of the centerings of the Gaussians from the 
lattice $m^*\Z$. The unperturbed potential thus 
takes its minima for $m\in m^*\Z$, 
the fixed parameter $m^*>0$ being its period.  
Later it will have to be large enough. 
(Note that the curvature of the potential
is of the order unity for large $m^*$; 
thus the curvature really has to be large on the rescaled
lattice where the potential has period $1$.)

We will simply take the 
$\left(\eta_x(h)\right)_{x\in \Z^d;h\in \Z}$ as i.i.d.
random variables with distribution $\P_{\eta}$
and the $\left(d_x(h)\right)_{x\in \Z^d;h\in \Z}$ as i.i.d.
random variables with  
distribution $\P_{d}$, independent of the $\eta$'s.
Furthermore we impose the smallness conditions

\item{(i)} $\P\left[|\eta_x(h)|\geq t\right]\leq e^{-\frac{t^2}{2\s^2_\eta}}$,
$\phantom{12345}$ (ii) $\phantom{1}|\eta_x(h)|\leq \d_{\eta}$

\item{(iii)} $\P\left[d_x(h)^2\geq t\right]\leq e^{-\frac{t^2}{2\s^2_d}}$,
$\phantom{12345}$ (iv) $\phantom{1}|d_x(h)|\leq \d_{d}\leq \frac{1}{4}$

\noindent where $\s^2_{d},\s^2_{\eta}\geq 0$ will be sufficiently small.

An assumption of the type (iv) is natural, since
it just states that the shifted wells stay away from each 
other and don't merge or even cross. 
The assumption (iii) is less natural
(and not really essential). 
Moreover we will need in the proof 
that  $\d_{d},\d_{\eta}$ be sufficiently small,
which is just to simplify the structure of the contour 
representation we will derive later and could be bypassed, see below.


To make explicit
the local dependence of various
quantities on the disorder variables
we write $\o_x=\left(d_x(h),\eta_x(h)\right)_{h\in \Z}$
for the `disorder variables at site $x$'  
and put $\o_{\L}=\left(\o_x \right)_{x\in \L}$.

A more general setting could of course be to consider
$V_x$ that are stationary w.r.t. a discrete shift 
in the height-direction
and satisfy some mixing condition. Also the i.i.d.
assumption in $x$ could be weakened.  




We use the following notation for 
the objects of interest, the 
finite volume Gibbs-measures $\mu_{\L}^{\tilde m_{\del \L},\o_{\L}}$,
defined in terms of their expectations:
$$
\eqalign{
&\mu_{\L}^{\tilde m_{\del \L},\o_{\L}}(f)
=\frac{1}{Z_{\L}^{\tilde m_{\del \L},\o_{\L}}}
\int_{\R^{\L}}dm_{\L}f\left(m_{\L},\tilde m_{\L^c} \right)
e^{-E^{\tilde m_{\del \L},\o_{\L}}_{\L}
\left(m_{\L}\right)}\text{where}\cr
&Z_{\L}^{\tilde m_{\del \L},\o_{\L}}
=\int_{\R^{\L}}dm_{\L}
e^{-E^{\tilde m_{\del \L},\o_{\L}}_{\L}
\left(m_{\L}\right)}\cr
}
\tag{1.3}
$$
for any bounded continuous $f$ on $\R^{\Z^d}$ (continuity is meant
w.r.t.  product topology).
Most of the times we will put zero boundary conditions
$\tilde m_x=0$ (for all $x\in \Z^d$), 
writing simply 
$\mu_{\L}^{0, \o_{\L}}$.
(Due to stationarity that's the same 
as putting $\tilde m_x=m^* l$ for any fixed $l\in \Z$.)

Then we have the stability result

\theo{1}{\it 
Let $d\geq 3$ and assume the conditions (i)-(iv) on the 
disorder variables.
Then, there exist $q_0>0$ (small enough), 
$\d_0>0$ (small enough),  
$\t_0<\infty$ (large enough), $\s_0^2>0$ (small enough)
such that, whenever $\d_{\eta},\d_{d}^2\leq \d_0$,
$q(m^*)^2\geq \t_0$, and $\s^{2}_{\hbox{\srm eff.}}:=
\s^2_{d}+\s_{\eta}^2\leq \s_0^2$,
the following is true. 

There exists an infinite volume random Gibbs-measure
$\mu^{\o}$ that can be obtained
as the weak limit 
$\mu^{\o}=\lim_{N\uparrow\infty}\mu^{0,\o_{\L_N}}_{\L_N}$
along a non-random sequence of cubes $\L_N$.
The measure describes a continuous interface 
localized around the base plane; its
`roughness' is bounded by 
$$
\eqalign{
&\E
\mu^{\o}
\left(m_{x_0}^2\right)
\leq 1+ (m^*)^2\left(e^{-\const \tilde \b}
+e^{-\frac{1}{\s^{\k}_{\hbox{\srm eff.}}}}\right)
\cr
}
\tag{1.4}
$$
for any $x_0$, with 
$\tilde \b=\const \times \min\left\{
\log\frac{1}{q}, q{m^*}^2 \left(\frac{\log \frac{1}{q}}{\log m^*}
\right)^d\right\}$ and an exponent $\k>0$.
}


\noindent{\thbf Remark: } 
So, measured on the scale of the period $m^*$, the roughness
is in fact a very small number. 
The term $1$ has to be present since
it describes the true fluctuations of a continuous spin 
in an individual well of $V_x$. 
The quantity $q (m^*)^2$ 
gives the true order of magnitude of the minimal 
energetic contribution of a pair of nearest neighbor heights
in neighboring potential wells, 
so it can be viewed as some basic temperature variable 
that has to be large enough. 
It appears in the definition of $\tilde \b$, with some
minor logarithmic deterioration that we need for technical
reasons. The $\log\frac{1}{q}$-contribution comes from
a high-temperature expansion, to be explained later.


Essential for the analysis 
is the following local transition kernel 
$T_x\left(\, \cdot\,\bigl| \,\cdot\,\right)$ describing
a single site coarse graining from 
a continuous height $m_x\in \R$ to 
an integer height $h_x\in \Z$.
It is defined by  keeping the Gaussian for $l=h_x$,
$$
\eqalign{
&T_x^{\o_x}\left(
h_{x}\bigl |m_x\right)
=\frac{e^{-\frac{1}{2}\left(m_x-m^*_x(h_x)\right)^2+\eta_x(h_x)}}
{\hbox{Norm.}(m_x)}, 
}
\tag{1.5}
$$
where $\hbox{Norm.}(m_x)$ is chosen to make
$T_x^{\o_x}\left(
h_{x}\bigl |m_x\right)$ a probability measure on $\Z$, for fixed $m_x$.
Note that this object depends on the disorder through $\o_x$
(as opposed to our proceeding in the double-well case
where a simpler analogous kernel was non-random).
So, for fixed $m_x$, the random probability 
weights $\left(T_x\left(
h\bigl |m_x\right)\right)_{h\in \Z}$ are integer
samples of a randomly perturbed Gaussian. 
The reader may also 
note the complementary fact that, for fixed $h_x$, the probability  
$T_x\left(h_{x}\bigl |m_x\right)$ is bounded 
by Gaussians from below and above and
has a unique absolute maximum as a function of $m_x$. The 
maximizer is close to $m^* h_x$.
[See Lemma 2.1 for these statements]. 
Now, the simplicity of our basic log-sum-of-Gaussian potential
(1.2) lies in the fact 
that $\hbox{Norm.}(m_x)=e^{-V_x(m_x)}$! 
[For more general potentials $V_x(m_x)$ 
this equality will acquire error terms (`anharmonicities')
that lead to additional expansions.]
We use the same symbol, 
$T\left(d h_{\L}\bigl| m_{\L} \right)
=\prod_{x\in \L}T^{\o_x}_x(h_x\bigl| m_x)$,
for the kernel from $\R^{\L}$
to $\Z^{\L}$, and also in the infinite volume.

Before we put down more results in a precise way, 
let us describe in an informal way in symbolic notation
what we are about to do.
Starting from a (finite volume) Gibbs measure $\mu(dm)$ 
we look at the joint distribution
$M(dh,dm):=T(dh|m)\mu(dm)$ on integer heights configurations
$h$ and continuous heights $m$. 
Then we analyse $\mu$ with the use of Bayes' formula:
We have $\mu(dm):=\int\nu(dh)  M(dm|h)$ where 
$\nu(dh)=\int\mu(dm) T(dh|m)$ is the $h$-marginal. 
It is of course a completely general (and a-priori empty) idea 
to look at distributions in suitably 
extended space that can only be useful 
for natural choices of this space. 
In our case we succeed with the control 
of $\nu(dh)$ since we can obtain a contour representation 
that can be treated by the spatial renormalization group. 
The conditional probability $M(dm|h)$ is nice
for the specific choice of the potential; 
it is just a Gaussian distribution. 
($M(dm|h)$ would be more complicated for perturbations of the 
potential, but the above decomposition would still be a successful one.)
Our results, to be described below,
will then concern the approach of the thermodynamic
limit of $\nu$. We will also have to clarify the 
interplay of the thermodynamic limit
with the above formulas.



The following theorem describes 
how the control of Gibbs-measures on the
integer heights carries over to the control
of the Gibbs-measures on the continuous heights,
under some harmless additional condition.

\theo{2}{\it  
Suppose that the discrete height measures 
$\nu^{\o}_{\L}:=
T\left(\mu_{\L}^{\tilde m_{\del\L},\o_{\L}}\right)$ 
converge locally along a sequence of 
cubes $\L_N$, centered at the origin,  to a limiting 
measure $\nu^{\o}(d h_{\Z^d})$
for a sequence of boundary conditions which is 
uniformly bounded, i.e. we have 
$\sup_{x\in \Z^d; N}
\left|\tilde m_{\del \L_N;x}\right|\leq M$, for some $M<\infty$. 
(That is, convergence
takes place for expectations of 
all bounded local observables.)
Assume moreover that we have the site-wise summability
$$
\eqalign{
&\sup_{N}\sum_{y\in \L_N}
\left(1-q\D_{\Z^d} \right)^{-1}_{x,y}
\nu^{\o}_{\L_N}|h_y|=:K_x(\o)<\infty
}
\tag{1.6}
$$
for a sequence of increasing cubes $\L_N$, for all $x\in \Z^d$
for a.e. configurations of the disorder $\o$.
Then the measures $\mu_{\L}^{\tilde m_{\del\L},\o_{\L}}$
converge locally to the infinite linear combination 
of Gauss measures given by 
$$
\eqalign{
&\mu^{\o}:=\int \nu^{\o}(d h_{\Z^d})
\NN\left[\left(1-q\D_{\Z^d} \right)^{-1}m^*_{\Z^d}
\left(h_{\Z^d} \right)
;
\left(1-q\D_{\Z^d}\right)^{-1}
\right]
}
\tag{1.7}
$$
That is,  convergence takes place
for sequences of 
expectations of bounded measurable $f(m_{V})$
that depend only on spins in the finite volume $V$.
}

The symbol 
$\NN\left[a;
\left(1-q\D_{\Z^d}\right)^{-1}\right]$ denotes
the massive Gaussian field on the infinite 
lattice $\Z^d$,
centered at $a\in \R^{\Z^d}$ with covariance matrix given by 
the second argument  (so that we have  e.g. $\int\NN\left[a;
\left(1-q\D_{\Z^d}\right)^{-1}\right](dm)(m_x-a_x)(m_y-a_y)=
\left[\left(1-q\D_{\Z^d}\right)^{-1}\right]_{x,y}$).

\remark Note that  the random 
quantities $K_x(\o)$ will typically not
be bounded uniformly in $x$. In fact, even a localized
interface will have unbounded fluctuations 
around regions of exceptionally large fluctuations
when considered  in the infinite lattice. 
Of course, (1.6) is implied by $\sup_{N,y}\E\nu^{\o}_{\L}|h_y|<\infty$.

\remark We stress that Theorem II does not only
apply to the `flat' interfaces that we investigate
here but also to more `exotic' Gibbs-measures. 
So, e.g. the (supposed) 
existence of Dobrushin-type integer-height 
Gibbs-measures (that are perturbations
of a flat interface at height $0$ in one 
half-space and a flat interface at height $H$ in the complement)
would imply the existence of corresponding 
continuous spin Gibbs-measures.

\bigskip
The organization of the paper is as follows.
In Chapter II we prove the `factorization' Theorem 2,
starting from its finite volume version Lemma 2.1.
In Chapter III we derive the contour representation 
of the integer height model (see Proposition 1),
starting from the finite volume Hamiltonian (3.1).
In Chapter IV we conclude to prove Theorem I 
from these results applying the spatial renormalization
group construction from [BoK1], [K1] on the contour model
representation.

\bigskip\bigskip

\chap{II. The Joint distributions 
of continuous and  integer heights\hfill}


Before we get started, let us make explicit some (concentration-)
properties of the random transition kernel 
to get some intuition for it. 
The elementary proof is 
given at the end of the chapter.

\lemma{2.1}{\it For any realization
of the disorder satisfying the bounds (ii) and (iv) below (1.2) 
the fixed-$h_x$ (random) probability   
$m_x\rightarrow T_x^{\o_x}\left(
h_{x}\bigl |m_x\right)
=\frac{e^{-\frac{1}{2}\left(m_x-m^*_x(h_x)\right)^2+\eta_x(h_x)}}
{\sum_{l\in \Z}e^{-\frac{1}{2}\left(m_x-m^*_x(l)\right)^2+\eta_x(l)}}$
has a unique absolute maximum. It has no other 
local maxima. 
The maximizer lies in the symmetric interval
about $m^* h_x$  with radius 
$\frac{m^*\left[4 \d_{d}+\d_d^2\right]}{4(1-\d_{d})}+\frac{1}{m^*2(1-\d_{d})}
\Biggl\{
\log\left[\frac{1+2\d_{d}}{1-2\d_{d}}\right]
+2\d_{\eta}
\Biggr\}$. We have that 
$$
\eqalign{
&\const e^{
-\frac{1}{2}
\left(m_x-m^*_x(h_x)  \right)^2}\leq 
T_x\left(
h_{x}\bigl |m_x\right)
\leq \Const e^{
-\frac{1}{2}
\left(m_x-m^*_x(h_x)  \right)^2}
}
\tag{2.1}
$$
with $\const>0$, $\Const>0$ depending only on $a$, $m^*$, $\d_{d}$, $\d_{\eta}$.
}

\medskip 

Now, the simplicity of our choice of the log-sum-of-Gaussian potential
lies in the fact that the joint distribution
on continuous heights and integer height 
can be written in the form
$$
\eqalign{
&\mu_{\L}^{\tilde m_{\del \L},\o_{\L}}(dm_{\L})
\prod_{x\in \L}T_x(h_x\bigl| m_x)
=\frac{1}{Z_{\L}^{\tilde m_{\del \L},\o_{\L}}}
e^{
-H^{\tilde m_{\del \L},\o_{\L},h_{\L}}_{\L}
\left(m_{\L}\right)}dm_{\L}
}
\tag{2.2}
$$
where 
$$
\eqalign{
&H^{\tilde m_{\del V},\o_{V}, h_{V}}_{V}
\left(m_{V}\right)\cr
&=\frac{q}{2}\sum_{{\{x,y\}\sb V}\atop {d(x,y)=1}}\left(
m_x-m_y
\right)^2+
\frac{q}{2}\sum_{{x\in V; y\in \del V}
\atop{d(x,y)=1}}\left(
m_x-\tilde m_y
\right)^2 + \frac{1}{2}\sum_{x\in V}\left(m_x-m^*_x(h_x)\right)^2 
-\sum_{x\in V}\eta_x(h_x)
}
\tag{2.3}
$$
is quadratic in $m$, for fixed $h$.
This is due to the cancellation of the normalization 
in the transition kernel against the exponential of the
potential.
We remark that, for potentials that can be viewed
as perturbations of 
our specific log-sum-of-Gaussians the formula would acquire 
error terms and the present formula is the main contribution
of a further expansion.  

We will now rewrite the joint distribution
as a product of the marginal in the integer heights
and the conditional distribution of the continuous heights
given the $h$. 
We see that the $m$-distribution conditioned
on a fixed value of $h$ is Gaussian.  
The $h$-marginals on the other hand can be computed 
by a Gaussian integration over $m_{\L}$:
Since the quadratic terms  of the above integral
are $h$-independent this Gaussian integration yields
$$
\eqalign{
&\int_{\R^{\L}}dm_{\L}e^{
-H^{\tilde m_{\del \L},\o_{\L},h_{\L}}_{\L}
\left(m_{\L}\right)}
=C_{\L}\times e^{
-\inf_{m_{\L}\in \R^{\L}}H^{\tilde m_{\del \L},\o_{\L},h_{\L}}_{\L}
\left(m_{\L}\right)}
}
\tag{2.4}
$$
with a constant $C_{\L}$ that does not depend on 
$h_{\L}$ (and $\o_{\L}$). 


By multiplying and dividing 
the r.h.s. of (2.1) by (2.3) we get after a little rewriting
of the Gaussian density, conditional on the $h$:

\lemma{2.2}{\it
The finite volume joint distribution 
of continuous and integer heights can be written as
$$
\eqalign{
&\mu_{\L}^{\tilde m_{\del \L},\o_{\L}}(dm_{\L})
\prod_{x\in \L}T_x(h_x\bigl| m_x)
=\nu^{\o_{\L}}_{\L}(h_{\L})\,\,
\NN\left[m_{\L}^{\tilde m_{\del \L},
\o_{\L},h_{\L}};\left(1-q\D^{D}_{\L}\right)^{-1}
\right](dm_{\L})\text{where}\cr
&\nu^{\o_{\L}}_{\L}(h_{\L})
=T\left(\mu_{\L}^{\tilde m_{\del\L},\o_{\L}}\right)(h_{\L})
=\frac{ e^{
-\inf_{m_{\L}\in \R^{\L}}H^{\tilde m_{\del \L},\o_{\L},h_{\L}}_{\L}
\left(m_{\L}\right)}}{\sum_{\tilde h_{\L}} e^{
-\inf_{m_{\L}\in \R^{\L}}H^{\tilde m_{\del \L},\o_{\L},\tilde 
h_{\L}}_{\L} \left(m_{\L}\right)}}\cr
}
\tag{2.5}
$$
with the random centering 
$$
\eqalign{
&m_{\L}^{\tilde m_{\del \L},
\o_{\L},h_{\L}}
=\left(1-q\D_{\L}   \right)^{-1}
\left(
m^*_{\L}(h_{\L})
+q\del_{\L,\del \L}\tilde m_{\del \L}
\right)
}
\tag{2.6}
$$
}

Here $\del_{\L,\del \L}$ is the matrix having entries 
$(\del_{\L,\del \L})_{x,y}=1$ iff $x\in \L$ and $y\in \del\L$
are nearest neighbors and zero otherwise. 
$\D_{\L}$ is the Lattice Laplacian (with Dirichlet boundary condition) 
in the volume $\L$.

{\bf Side-remark:} We like to point out 
that the enlargement of the probability 
space by the introduction of auxiliary 
integration variables and conditioning on the 
latter ones can be found in various
places in statistical mechanics: 

1) In the renormalization group analysis
one studies the Gibbs-distributions $\mu(\G)$
of the variables $\G$ of the system with the aid
of a mapping to spatially coarse-grained variables $\G'$
by means of a transition kernel $T(d\G'|\G)$. This gives
rise to a joint distribution $M(d\G,d\G')=\mu(d\G)T(d\G'|\G)$.
Reversing the order of conditioning gives 
$M(d\G,d\G')=\nu(\G')M(d\G|\G')$, the idea being that 
the `renormalized' measures $\nu(\G')$ are easier to study than 
the measures $\mu(\G)$. (Of coarse, even if this is true,
the conditional distribution  $M(d\G|\G')$ must also 
be controlled.)  

2) The introduction 
of artificial integration variables  is a commonly used trick 
also for the analysis in quadratic mean-field models
(known here as Hubbard-Stratonovitch
transformation). 
In fact, the analogue of  formulas (2.5) 
looks as follows for the simplest candidate, 
the usual mean-field Ising ferromagnet.  
Its Gibbs distribution on the spins $(\s_i)_{i=1,\dots,N}\in
\{-1,1\}^N=:\O$ is given by 
$\mu_N(\s)= e^{\frac{\b N}{2}m_N(\s)^2}/\hbox{Norm.}$
where $m_N(\s)=\frac{1}{N}\sum_{i=1}^N\s_i$.
Consider the variables $(\s,\tilde m)$ in
the enlarged probability space $\O\times \R$ (with
a new smoothed out magnetization variable $\tilde m$) 
that are distributed according to the joint distribution 
$M(\s,d\tilde m)=
e^{\b N m(\s)\cdot \bar m}e^{-\frac{\b N}{2}\tilde m^2}
d\tilde m/\hbox{Norm.}$. Then the marginal distribution 
on the $\s$ is the desired Gibbs-distribution;
in fact we have 
$M(\s,d\tilde m)=\mu(\s)T(\tilde m|\s)$
with $T(d\tilde m|\s)=e^{-\frac{\b N}{2}(\tilde m-m(\s))^2}d\tilde m
/\hbox{Norm.}$ describing the smoothed out averaging
over the whole lattice. 
To analyse the Gibbs-distribution the
joint distribution is then written by 
reversing the order of conditioning in the form 
$M(\s,d\tilde m)=\nu(d\tilde m)M(\s|\tilde m)$
where  $\nu(d\tilde m)$ is the 
marginal on the $\tilde m$ and $M(\s|\tilde m)=\prod_i 
M_i(\s_i|\tilde m)$ where
$M_i(\s_i|\tilde m)=e^{\b \tilde m \s_i-\log 2\cosh(\b\tilde m)}$.
The latter kernel is clearly trivial, the distribution 
$\nu(d\tilde m)$ is treated by a saddle-point method. 

Likewise, our strategy in the present problem
will now be to control the $\nu$-distribution
in the thermodynamic limit 
and get the Gibbs-measures of the continuous spins 
by summing (2.5) over $h$.
Assuming this control over the integer heights
we must however also control what happens to 
(the $h$-average over)
equation (2.5) under the thermodynamic limit 
if we apply it to a local function
$f(m_V)$, depending on continuous heights $m_x$
only for $x\in V$, $V$ being a fixed finite volume.
Note that the Gaussian describing the 
conditional distribution of the continuous
heights, given the integer heights, has some $\L$-
dependence both through its centering and the
covariance matrix. Further, its dependence 
on $h$ is not finite range (albeit strongly decaying
unless the integer-heights are getting very large.)
So we need some extra condition on the convergence 
of $\nu_{\L}$-measure and a little work 
to deduce the implication of the 
desired convergence of the $\mu_{\L}$-measure. 

The precise result of this is given in Theorem 2 (see Introduction), 
that we are going to prove now. While doing so, 
we will also prove the following 

{\bf Addition to Theorem 2:}
{\it Assume the site-wise  existence of all exponential moments
$$
\eqalign{
&\sup_{\L}\nu^{\o}_{\L}\left[e^{s\sum_{y\in \L}
\left(1-q\D_{\Z^d} \right)^{-1}_{x,y}
|h_y|}\right]=:\tilde K_x(\o,s)<\infty,
}
\tag{2.7}
$$
Then the  convergence 
$\lim_{\L\uparrow\infty}
\mu_{\L}^{\tilde m_{\del\L},\o_{\L}}(f_{V})=
\mu^{\o}(f_{V})$
takes place also  for all local observables $f(m_{V})$
that do not increase faster than exponentially;
i.e. there exists a constant $\l\geq 0$ s.t. 
$|f(m_{V})|\leq e^{\l\Vert m_{V}\Vert_2}$.}

{\thbf Proof of Theorem 2:}
We start with the proof of the theorem under the `site-wise
summability assumption' (1.5).
We must control large realizations of
the $h$'s (that are however improbable w.r.t 
$\nu$, under this assumption.)
Let $f$ denote any measurable function of $m_V$,
we assume for simplicity that $f$ is uniformly bounded by $1$.
To produce a local observable (of the integer heights) 
we cut off the long range dependence of the Gaussians
on the integer height $h$ outside some 
volume $\L_2$ that satisfies
$V\sb\L_2\sb\L$. We use an $\e/3$-trick to decompose  
$$
\eqalign{
&\left|\nu_{\L}^\o\NN\left[m_{\L}^{\tilde m_{\del \L},
\o_{\L},h_{\L}};\left(1-q\D^{D}_{\L}\right)^{-1}
\right](f)
-\nu_{\infty}^\o\NN\left[\left(1-q\D_{\Z^d} \right)^{-1}m^*_{\Z^d}
\left(h_{\Z^d} \right)
;
\left(1-q\D_{\Z^d}\right)^{-1}
\right](f)\right|
\cr
&\leq 
\nu_{\L}^\o\Biggl|
\NN\left[m_{\L}^{\tilde m_{\del \L},
\o_{\L},h_{\L}};\left(1-q\D^{D}_{\L}\right)^{-1}
\right](f)\cr
&\qquad-\NN\left[
\left(1-q\D_{\Z^d} \right)^{-1}\left(m^*_{\L_2}
\left(h_{\L_2} \right),
0_{\L_2^c}
\right);\left(1-q\D_{\Z^d}\right)^{-1}
\right](f)
\Biggr|\cr
&+\left|
\left(\nu_{\L}^\o-\nu_{\infty}^\o\right)
\NN\left[
\left(1-q\D_{\Z^d} \right)^{-1}\left(m^*_{\L_2}
\left(h_{\L_2} \right),
0_{\L_2^c}
\right);\left(1-q\D_{\Z^d}\right)^{-1}
\right](f)\right|\cr
&+
\nu_{\infty}^\o\Biggl|
\NN\left[\left(1-q\D_{\Z^d} \right)^{-1}m^*_{\Z^d}
\left(h_{\Z^d} \right)
;
\left(1-q\D_{\Z^d}\right)^{-1}
\right](f)\cr
&\quad-\NN\left[
\left(1-q\D_{\Z^d} \right)^{-1}\left(m^*_{\L_2}
\left(h_{\L_2} \right),
0_{\L_2^c}
\right);\left(1-q\D_{\Z^d}\right)^{-1}
\right](f)
\Biggr|\cr
}
\tag{2.8}
$$
We will show that the r.h.s. can be made
arbitrarily small be choosing at first the auxiliary $\L_2$
and then $\L$ large enough. 
Indeed, the middle term on the r.h.s. converges to zero
with $\L\uparrow\infty$, for any fixed $\L_2$, due to the assumption
of weak convergence of the $\nu^\o_\L$. 
The remaining task is to control the two error-terms;
for this we need the condition (1.5) [resp. (2.7)]. 

We look at the first term on the r.h.s.
more carefully. The last term is treated in a similar fashion. 
We need some continuity properties 
of $|V|$-dimensional Gaussian expectations
considered as functions of their means and covariances.
The following estimate
will do, both for bounded observables
and observables that are only exponentially bounded.

\lemma{2.3}{\it Let $\NN[a,\S]$,
$\NN[a',\S']$ denote
two $|V|$-dimensional non-degenerate
Gaussians with mean $a,a'\in \R^{V}$ 
and covariance $\S,\S'\in \R^{V\times V}$. 
Assume that $f(m_V)$ is an observable that doesn't
increase faster than exponentially, i.e.
$|f(m_V)|\leq e^{\l\Vert m_V \Vert_2}$ for some $\l\geq 0$.
Then we have the following estimate
$$
\eqalign{
&\left|\int \NN\left[a;\S\right](d m_V)f(m_V)
- \int \NN\left[a';\S'\right](d m_V)f(m_V)\right|\cr
&\leq  2^{|V|}e^{\l\Vert a\Vert_2}e^{\frac{\l^2 \hbox{\srm Tr}\S}{2}}\Biggl[\left|
1-\left(\frac{\det \S}{\det \S'}\right)^{\frac{1}{2}}
\right|\cr 
&+\left(\frac{\det \S}{\det \S'}\right)^{\frac{1}{2}}
g\left(
\left( 2 S 
+\Vert a\Vert_2 +\Vert a'\Vert_2
\right)\Vert{\S}^{-1}\Vert_2 \,    \Vert a- a'\Vert_2
+2 \left(S^2+  \Vert a'\Vert_2^2 \right)
\times\Vert{\S}^{-1}-{\S'}^{-1}\Vert_2
\right)\Biggr]\cr
&+2^{|V|}e^{\l S
-\frac{(S-\Vert a\Vert_2)^2 }{2\hbox{\srm Tr}\S}}
+2^{|V|}e^{\l S
-\frac{(S-\Vert a'\Vert_2)^2 }{2\hbox{\srm Tr}\S'}}
}
\tag{2.9}
$$
where $g(x)=x e^{x}$, for any 
$S\geq  \max\{\Vert a\Vert_2+\l\hbox{Tr}\S,
\Vert a'\Vert_2+\l\hbox{Tr}\S'\}$.
}

\proof To show the Lemma we 
decompose the range of integration into 
a ball of radius $S$ and its complement;
to control 
the corresponding integrals we need a simple application of the exponential
Markov inequality in the form of 

\medskip

\lemma{2.4}{\it Let $\NN[a,\S]$ denote
a Gaussian with mean $a\in \R^{V}$ 
and covariance $\S\in \R^{V\times V}$ (i.e.
$\int\NN[a,\S](dm_V)\left(m_x-a_x\right)\left(m_y-a_y\right)=\S_{x,y}$).
Then we have 
$$
\eqalign{
&\int \NN\left[a;\S\right](d m_V)
e^{\l\Vert m_V\Vert_2}
\leq  2^{|V|}e^{\l\Vert a\Vert_2}e^{\frac{\l^2 \hbox{\srm Tr}\S}{2}}\text{and}\cr
&\int \NN\left[a;\S\right](d m_V)
e^{\l\Vert m_V\Vert_2}1_{\Vert m_V\Vert_2\geq S}
\leq  2^{|V|}e^{\l S
-\frac{(S-\Vert a\Vert_2)^2 }{2\hbox{\srm Tr}\S}}\text{for}
S\geq \Vert a\Vert_2+\l\hbox{Tr}\S
}
\tag{2.10}
$$
For $S=\Vert a\Vert_2+\l\hbox{Tr}\S$
the two r.h.s. coincide.
}

\proof
Write $\int \NN\left[a;\S\right](d m_V)
e^{\l\Vert m_V\Vert_2}
\leq e^{\l\Vert a\Vert_2} 
\int \NN\left[a;\S\right](d m_V)
e^{\l\Vert m_V-a\Vert_2}$
and denote by $\s_i^2$ the eigenvalues of $\S$
and by $e_i$ the corresponding eigenvectors. 
Then we may write 
$$
\eqalign{
&\int \NN\left[a;\S\right](d m_V)
e^{\l\Vert m_V-a\Vert_2}
\leq 
\left(\prod_{i=1,\dots,|V|}\int \NN\left[0;\s_i^2\right](d\hat m_{i})
e^{\l\left| \hat m_{i}\right|}\right)
\leq  
2^{|V|}
e^{\frac{\l^2 \sum_{i}\s_i^2}{2}}\cr
}
\tag{2.11}
$$
which proves the first estimate. The second one
is a corollary: Write, for nonnegative $\l_1$,
$$
\eqalign{
&\int \NN\left[a;\S\right](d m_V)
e^{\l\Vert m_V\Vert_2}1_{\Vert m_V\Vert_2\geq S}
\leq  e^{-\l_1 S}\int \NN\left[a;\S\right](d m_V)
e^{(\l+\l_1)\Vert m_V\Vert_2}1_{\Vert m_V\Vert_2\geq S}\cr
&\leq  e^{-\l_1 S}2^{|V|}e^{(\l+\l_1)
\Vert a\Vert_2}e^{\frac{(\l+\l_1)^2 \hbox{\srm Tr}\S}{2}}
}
\tag{2.12}
$$
where the last inequality follows from the first claim.
Minimizing the r.h.s. yields the claim (the optimal
value of $\l_1$ being $\l_1=\frac{S-\Vert a\Vert_2}{\hbox{\srm Tr}\S}-\l$).
The requirement that $\l_1$ be positive leads to the given 
range of allowed $S$.
\endproof
\medskip

We continue with the proof of the Lemma 2.3 writing 
$$
\eqalign{
&\left|\int \NN\left[a;\S\right](d m_V)f(m_V)
- \int \NN\left[a';\S'\right](d m_V)f(m_V)\right|\cr
&\leq 
\left|\left(\int \NN\left[a;\S\right]- \int \NN\left[a';\S'\right]\right)
(d m_V)f(m_V)1_{|m_{V}|\leq S}\right|\cr
&\quad+\int \NN\left[a;\S\right]e^{\l\Vert m_V \Vert}1_{|m_{V}|\geq S}
+\int \NN\left[a';\S'\right]e^{\l\Vert m_V \Vert}1_{|m_{V}|\geq S}\cr
}
\tag{2.13}
$$
The last two terms are estimated with the help of the above lemma,
leading to the last line of (2.9).
The first term can be estimated simply in terms of differences
of the Gaussian  densities:
$$
\eqalign{
&\left|\int dm_{V}\left(\frac{e^{-\frac{1}{2}<(m_{V}-a),\S^{-1}(m_{V}-a)>}}
{(2\pi)^{\frac{|V|}{2}}(\det \S)^{\frac{1}{2}}}
-\frac{e^{-\frac{1}{2}<(m_{V}-a'),{\S'}^{-1}(m_{V}-a')>}}
{(2\pi)^{\frac{|V|}{2}}(\det \S')^{\frac{1}{2}}}
\right)f(m_V)1_{|m_{V}|\leq S}\right|\cr
&\leq \sup_{|m_{V}|\leq S}\left|
1-e^{\frac{1}{2}\left[
<(m_{V}-a),{\S}^{-1}(m_{V}-a)>
-<(m_{V}-a'),{\S'}^{-1}(m_{V}-a')>\right]}
\left(\frac{\det \S}{\det \S'}\right)^{\frac{1}{2}}
\right|\cr
&\qquad\times 
\int dm_{V}\frac{e^{-\frac{1}{2}<(m_{V}-a),\S^{-1}(m_{V}-a)>}}
{(2\pi)^{\frac{|V|}{2}}(\det \S)^{\frac{1}{2}}}f(m_V)1_{|m_{V}|\leq S}\cr
}
\tag{2.14}
$$
The last term is estimated by dropping the 
characteristic function and applying the first statement in Lemma 2.4.
The last $\sup$ is estimated from above by
$$
\eqalign{
&\left|
1-\left(\frac{\det \S}{\det \S'}\right)^{\frac{1}{2}}
\right|
+\left(\frac{\det \S}{\det \S'}\right)^{\frac{1}{2}}
\sup_{|m_{V}|\leq S}
\left|
1-e^{\frac{1}{2}\left[
<(m_{V}-a),{\S}^{-1}(m_{V}-a)>
-<(m_{V}-a'),{\S'}^{-1}(m_{V}-a')>\right]}\right|\cr
}
\tag{2.15}
$$
where, using the simple estimate $|e^x-1|\leq |x|e^{|x|}=:g(|x|)$
the sup in the last expression can be estimated by 
$
g\left[\frac{1}{2}
\sup_{|m_{V}|\leq S}
\left|
<(m_{V}-a),{\S}^{-1}(m_{V}-a)>
-<(m_{V}-a'),{\S'}^{-1}(m_{V}-a')>\right|
\right]$.
For this last sup in the argument of $g$ we use the upper
estimates in terms of two-norms
$$
\eqalign{
&2\sup_{|m_{V}|\leq S}
\left|<m_{V},{\S}^{-1}(a-a')>\right|
+\left|<a,{\S}^{-1}a>- <a',{\S}^{-1}a'>\right|
\cr
&\qquad+\sup_{|m_{V}|\leq S}
\left|
<(m_{V}-a'),\left[{\S}^{-1}-{\S'}^{-1} \right](m_{V}-a')>\right|\cr
&\leq 
\left( 2 S 
+\Vert a\Vert_2 +\Vert a'\Vert_2
\right)\Vert{\S}^{-1}\Vert_2 \,    \Vert a- a'\Vert_2
+2 \left(S^2+  \Vert a'\Vert_2^2 \right)
\times\Vert{\S}^{-1}-{\S'}^{-1}\Vert_2
\cr
}
\tag{2.16}
$$
Collecting our results gives Lemma 2.3.
\endproof
\medskip

To apply the Lemma we just use the short notation 
$$
\eqalign{
&a:= a(\L):=\Pi_V m_{\L}^{\tilde m_{\del \L},
\o_{\L},h_{\L}}
=\Pi_V\left(1-q\D_{\L}   \right)^{-1}
\left(
m^*_{\L}(h_{\L})
+q\del_{\L,\del \L}\tilde m_{\del \L}
\right)\cr
&a':=a'(\L_2):=\Pi_V\left(1-q\D_{\Z^d} \right)^{-1}\left(m^*_{\L_2}
\left(h_{\L_2} \right),
0_{\Z^d\ba \L_2}
\right)\cr
}
\tag{2.17}
$$
for the expectations of the $|V|$-dimensional Gaussians
under consideration. 
We also denote by $\S:=\Pi_V\left(1-q\D_{Z^d}\right)^{-1}\Pi_V$ 
the infinite volume covariance matrix restricted 
to $V$, and correspondingly 
$\S':=\S'(\L):=\Pi_V\left(1-q\D^{D}_{\L}\right)^{-1}\Pi_V$.

The volume difference of the covariances is fairly harmless:
Given $\e>0$ we can choose
$\L_0$ sufficiently large, s.t. for all $\L\sp \L_0$,
we have that $\left|
1-\left(\frac{\det \S}{\det \S'(\L)}\right)^{\frac{1}{2}}
\right|\leq \e$ and $\Vert{\S}^{-1}-{\S'(\L)}^{-1}\Vert_2\leq \e$.
Further, all matrix elements of $\S'(\L)$
are bounded from above by the 
corresponding infinite volume expression $\S$.
In particular, we can use the upper bound
$\hbox{Tr}\S'(\L)\leq \hbox{Tr}\S$.
(All of this can be explicitly seen from the 
random walk representation of the resolvent, see e.g. [K4])

Assuming these choices we get with Lemma 2.3
$$
\eqalign{
&\left|\int \NN\left[a;\S\right](d m_V)f(m_V)
- \int \NN\left[a';\S'\right](d m_V)f(m_V)\right|\cr
&\leq  2^{|V|}
\Biggl[\e
+(1+\e)
g\left(
\left( 2 S 
+\Vert a\Vert_2 +\Vert a'\Vert_2
\right)\Vert{\S}^{-1}\Vert_2 \,    \Vert a- a'\Vert_2
+2\e \left(S^2+  \Vert a'\Vert_2^2 \right)\right)\Biggr]\cr
&+2^{|V|}e^{-\frac{(S-\Vert a\Vert_2)^2 }{2\hbox{\srm Tr}\S}}
+2^{|V|}e^{
-\frac{(S-\Vert a'\Vert_2)^2 }{2\hbox{\srm Tr}\S}}
}
\tag{2.18}
$$
for $\L\sp \L_0(\e)$
where the $\S$-terms appear now as fixed constants.

To estimate the  $\nu_{\L}^\o$-expectation of this bound
we will decompose the space of the integer heights into a `regular set' 
$\HH:=\HH(\L_2,\L):=\HH^{(1)}(\L_2,\L)\cap \HH^{(2)}(\L_2,\L)$
where 
$$
\eqalign{
&\HH^{(1)}(\L_2,\L):=\left\{h_{\Z^d}; \Vert a(\L)\Vert_2\leq B, 
\Vert a'(\L_2)\Vert_2\leq B\right\}\cr
&\HH^{(2)}(\L_2,\L):=\left\{h_{\Z^d};
  \Vert a(\L)-a'(\L_2)\Vert_2\leq \e_2
\right\}\cr
}
\tag{2.19}
$$
and its complement. We get from this (with $\Vert f\Vert_{\infty}\leq 1$) that 
$$
\eqalign{
&\nu_{\L}^\o\Biggl|
\NN\left[m_{\L}^{\tilde m_{\del \L},
\o_{\L},h_{\L}};\left(1-q\D^{D}_{\L}\right)^{-1}
\right](f)
-\NN\left[
\left(1-q\D_{\Z^d} \right)^{-1}\left(m^*_{\L_2}
\left(h_{\L_2} \right),
0_{\L_2^c}
\right);\left(1-q\D_{\Z^d}\right)^{-1}
\right](f)
\Biggr|\cr
&\leq \nu_{\L}^\o\left[{\HH(\L_2,\L)}^c \right]
+ 2^{|V|}
\Biggl[\e
+(1+\e)
g\left(
\left(2 S 
+B + B
\right)\Vert{\S}^{-1}\Vert_2 \, \e_2
+2\e \left(S^2+ B^2 \right)
\right)\Biggr]\cr
&+2^{|V|}e^{-\frac{(S-B)^2 }{2\hbox{\srm Tr}\S}}
+2^{|V|}e^{
-\frac{(S-B)^2 }{2\hbox{\srm Tr}\S}}
}
\tag{2.20}
$$
To estimate the exceptional set of integer heights
we will prove below 

\lemma{2.5}{\it 
\item{(i)} For all (arbitrarily small)
$\d$ there exists a $B<\infty$ (sufficiently large)
s.t.
$\nu_{\L}^\o\left[{\HH^{(1)}(\L_2,\L)}^c \right]\leq \d$
for all sufficiently large $\L_2,\L$.

\item{(ii)} For all (arbitrarily small)
$\d,\e_2$, there exist 
choices of volumes $\L_2\sb\tilde\L_0$ (suff. large) 
s.t. $\phantom{1234}\nu_{\L}^\o\left[{\HH^{2}(\L_2,\L)}^c \right]\leq \d$
whenever $\L\sp \tilde \L_0$.
}

\medskip

But assuming this property, (2.20)
can be made smaller than any given $\d$ for sufficiently large $\L$
in the following way:

1) Choose $B=S/2$ large enough that a) the sum of the last two terms 
is smaller than $\d/3$ and b)
$\nu_{\L}^\o\left[{\HH^{(1)}(\L_2,\L)}^c \right]\leq \d/6$,
according to Lemma 2.5(i).
(So we must have that both $\L_2,\L$ are large enough.)

2) Given these choices  of $S,B$, choose $\e,\e_2$ small enough s.t.
the middle term is smaller than $\d/3$. (Then the estimates
hold true, after possibly enlarging $\L$.)

3) Finally there are then choices  of $\L_2\sb\tilde\L_0$ 
s.t. 
$\nu_{\L}^\o\left[{\HH^{(2)}(\L_2,\L)}^c \right]\leq \d/6$
for all $\L\sp \tilde\L_0$, according to Lemma 2.5(ii).

This finishes our discussion of the proof of Theorem 2; 
it remains however to give the {\thbf Proof of Lemma 2.5:}
In fact, the Lemma holds 
under the following two weaker conditions of 

\item{(a)} Uniform integrability
$\lim_{B\uparrow\infty}\sup_{\L}\nu^{\o}_{\L}\left[
\sum_{y\in \Z^d}R_{\infty;x,y}|h_y|\geq B
\right]=0$

\item{(b)} Uniform summability 
$\lim_{R\uparrow\infty}\sup_{\L}\nu^{\o}_{\L}\left[
\sum_{y\in \Z^d; |y|\geq R}R_{\infty;x,y}|h_y|\geq \e
\right]=0$

for each $x\in \Z^d$.
(These condition are 
implied from the hypothesis by 
Chebycheff, e.g. 
$\sup_{\L}\nu^{\o}_{\L}\left[
\sum_{y\in \Z^d}R_{\infty;x,y}|h_y|\geq B
\right]\leq \frac{1}{B}\sum_{y\in \Z^d}R_{\infty;x,y}
\sup_{\L}\nu^{\o}_{\L}\left[
|h_y|\right]$.)

{\bf (i):}
Note that, due to the exponential decay of the resolvent
we have for uniformly bounded boundary conditions
that 
$\lim_{\L\uparrow\infty}\Pi_V\left(1-q\D_{\L}   \right)^{-1}
q\del_{\L,\del \L}\tilde m_{\del \L}=0$. 
It suffices to look at one matrix element, say $a_x(\L)$ of the 
vector $a(\L)$.
Using that $R_{\L;x,y}\leq R_{\infty;x,y}$
uniform in the volume,
the form of $m^*(h)$, and the uniform boundedness of 
the random shift in the continuous spin Hamiltonian, it is 
immediate to see that what we need is implied by 
the above uniform integrability condition.

{\bf (ii):}
The differences are estimated as follows. For $x\in V$ we have 
$$
\eqalign{
&\Biggl|\left[\left(1-q\D_{\L}   \right)^{-1}
\left(
m^*_{\L}(h_{\L})
+q\del_{\L,\del \L}\tilde m_{\del \L}
\right)\right]_x
-\left[\left(1-q\D_{\Z^d} \right)^{-1}\left(m^*_{\L_2}
\left(h_{\L_2} \right),
0_{\Z^d\ba \L_2}
\right)\right]_x\Biggr|\cr
&\leq\sum_{y\in \L_2}
\left| 
\left(1-q\D_{\L}   \right)^{-1}_{x,y}
-\left(1-q\D_{\Z^d} \right)^{-1}_{x,y}
\right| \left|m^*_{y}
\left(h_{y} \right)\right|\cr
&+\sum_{y\in \L\ba\L_2}
\left(1-q\D_{\L}   \right)^{-1}_{x,y}
\left|m^*_{y}\left(h_{y} \right)\right|
+q M\sum_{y\in \del\L}
\left(1-q\D_{\L}   \right)^{-1}_{x,y}
\cr
&\leq\sup_{y\in \L_2}\left|
\frac{\left(1-q\D_{\L}   \right)^{-1}_{x,y}}{\left(1-q\D_{\Z^d} \right)^{-1}_{x,y}}
-1\right|
 \sum_{y\in \L_2}
\left(1-q\D_{\Z^d} \right)^{-1}_{x,y} \left|m^*_{y}
\left(h_{y} \right)\right|\cr
&+\sum_{y\in \L\ba\L_2} 
\left(1-q\D_{\Z^d}   \right)^{-1}_{x,y}
\left|m^*_{y}\left(h_{y} \right)\right|
+q M\sum_{y\in \del\L}
\left(1-q\D_{\Z^d}   \right)^{-1}_{x,y}
\cr
}
\tag{2.21}
$$
We need to show that  the $\nu_{\L}^{\o}$-probability
of the event that the r.h.s. is bigger than some $\tilde\e_2$
can be made small by choosing the volumes in a useful way. 
The last (deterministic) term converges to zero;
so we assume that $\L$ is large enough s.t. it is smaller than
$\tilde \e_2/3$.  
Now, from (b) we know that, given any $\d$, we have 
for all sufficiently large $\L_2$ that 
$\sum_{y\in \L\ba\L_2} 
\left(1-q\D_{\Z^d}   \right)^{-1}_{x,y}
\left|m^*_{y}\left(h_{y} \right)\right|\leq \tilde \e_2/3
$
with (say) $\nu_{\L}^{\o}$-probability bigger than $1-\d/2$.
We fix such a $\L_2$.
What we have just seen in the proof of (i) 
ensures that, for given $\d$ we 
can find a $B$ such that the sum of $y\in \L_2$ on the r.h.s.
of the last inequality is bounded by $B$, uniformly in $\L_2$,
with (say) $\nu_{\L}^{\o}$-probability bigger than $1-\d/2$.
Now it remains to choose $\L$ is large as we want to make
the $\sup$ over $y$'s in the fixed $\L_2$ as small as we want,
and thus the first line on the r.h.s. smaller than $\tilde \e_2/3$
to finish the proof of Lemma 2.5.\endproof

\bigskip
Let us finally give the modifications
needed to get the 

\noindent{\bf Proof of the Addition to Theorem 2:}.
Let us look again at first at the 
first term of the decomposition (2.8) where $f$ is
now a local observable that is only exponentially bounded.
We introduce the same type of exceptional 
set $\HH(\L_2,\L)$. Then, after
using the Lemma 2.4 for the Gaussian
expectation {\it on} the exceptional set,
the analogue of (2.20) becomes 
$$
\eqalign{
&\nu_{\L}^\o\left|
\NN\left[m_{\L}^{\tilde m_{\del \L},
\o_{\L},h_{\L}};\left(1-q\D^{D}_{\L}\right)^{-1}
\right](f)
-\NN\left[
\left(1-q\D_{\Z^d} \right)^{-1}\left(m^*_{\L_2}
\left(h_{\L_2} \right),
0_{\L_2^c}
\right);\left(1-q\D_{\Z^d}\right)^{-1}
\right](f)
\right|\cr
&\leq 2^{|V|}e^{\frac{\l^2 \hbox{\srm Tr}\S}{2}}
\nu_{\L}^\o\left[e^{\l\Vert a\Vert_2}1_{{\HH(\L_2,\L)}^c} \right]
+ 2^{|V|}e^{\l B}e^{\frac{\l^2 \hbox{\srm Tr}\S}{2}}\cr
&\times \Biggl[\e
+(1+\e)
g\left(
\left(2 S 
+B + B
\right)\Vert{\S}^{-1}\Vert_2 \, \e_2
+2\e \left(S^2+ B^2 \right)
\right)\Biggr]+2^{|V|}e^{\l S-\frac{(S-B)^2 }{2\hbox{\srm Tr}\S}}
+2^{|V|}e^{\l S-\frac{(S-B)^2 }{2\hbox{\srm Tr}\S}}
}
\tag{2.22}
$$
To treat the $\nu$-integral over the exceptional set, 
use the Schwartz inequality
$\nu_{\L}^\o\left[e^{\l\Vert a\Vert_2}1_{{\HH(\L_2,\L)}^c} \right]
\leq \left(\nu_{\L}^\o\left[e^{2\l\Vert a\Vert_2}
\right]\right)^{\frac{1}{2}}
\nu_{\L}^\o\left[\HH(\L_2,\L)^c \right]^{\frac{1}{2}}
$.
Now, given the assumption of the existence
of exponential moments, the first term on the r.h.s.
can easily seen to be bounded by a constant independent of $\L$.
But after this, we are essentially 
in the same situation as after (2.20)
and the way of  choosing  
the parameters stays the same as before. 
\endproof 

\bigskip
We are still due the

{\thbf Proof of Lemma 2.1:} 
By joint shift in height-direction 
we can assume that $h_x=0$.
We write the kernel in the form
$T_x\left(
h_{x}=0\bigl |m_x\right)
=\left[1 +\sum_{h\in\Z;h\neq 0} f_h(m_x)
\right]^{-1}$
with 
$f_h(m_x):=e^{(m^*_x(h_x)-m^*_x(h_x=0))m-
\frac{1}{2}\left[m^*_x(h_x)^2-m^*_x(h_x=0)^2\right]
+\eta_x(h_x)-\eta_x(h_x=0)}$. 
To prove unicity of the local maximum of the kernel
we note  that $m_x\mapsto\sum_{h\in\Z;h\neq 0} f_h(m_x)$
(being a sum of strictly convex functions) is a strictly convex function; 
hence it has a unique
local minimum which is the global minimum.  

To prove the bounds on the minimizer we look
at the individual minimizers $\bar m(h)$ 
of each of the pairs 
$f_{h}+f_{-h}$ for $h=1,2,\dots$.
We will show that $|\bar m(h)|\leq A$,
for all $h$. This implies that the minimizer 
of $m_x\mapsto\sum_{h\in\Z;h\neq 0} f_h(m_x)$ 
satisfies the same bound (since all terms in the sum
are strictly decreasing [increasing]
for $m_x\geq A$ [$\leq -A$].) 
Now, a computation gives
$$
\eqalign{
&\bar m(h)
=\frac{1}{m^*\left(
2h +d_x(h)-d_x(-h)\right)}
\Biggl\{
\log\left[\frac{h-d_x(-h)+d_x(h=0)}{h+d_x(-h)-d_x(h=0)}\right]\cr
&+\frac{(m^*)^2}{2}\left[2 h(d_x(h)+d_x(-h))
+d_x(h)^2-d_x(-h)^2\right]
-\eta_x(h)+\eta_x(-h)
\Biggr\}
}
\tag{2.23}
$$
Substituting the a-priori bounds
of the random quantities $|d_x(h)|\leq \d_{d}$
and $|\eta_x(h)|\leq \d_{\eta}$ we get 
from this 
$$
\eqalign{
&\left|\bar m(h)\right|
\leq \frac{1}{m^*2h(1-\d_{d})}
\Biggl\{
\log\left[\frac{h+2\d_{d}}{h-2\d_{d}}\right]
+\frac{(m^*)^2}{2}\left[4 h\d_{d}
+\d_d^2\right]
+2\d_{\eta}
\Biggr\}\cr
&\leq \frac{m^*\left[4 \d_{d}
+\d_d^2\right]}{4(1-\d_{d})}
+\frac{1}{m^*2(1-\d_{d})}
\Biggl\{
\log\left[\frac{1+2\d_{d}}{1-2\d_{d}}\right]
+2\d_{\eta}
\Biggr\}\cr
}
\tag{2.24}
$$
To see the last estimates, 
just look at the nominator
$\hbox{Norm.}(m_x)$ (see 1.4):
It is simple to check that this sum converges and
it is  bounded from above as well
as bounded from below away from $0$. 
From this the bounds (2.1) follow. 
\endproof
\bigskip\bigskip

\chap{III. A useful contour representation for discrete heights\hfill}

In this chapter we will treat the measures $\nu^{\o}_{\L}$
for the discrete height model that 
are given by (2.5) with (2.3).
The effective finite volume Hamiltonian for the integer heights
can be simply computed as a minimum of a quadratic expression
in continuous variables: 
For any boundary condition $\tilde m$ it reads
$$
\eqalign{
&\inf_{m_{\L}\in \R^{\L}} H^{\tilde m_{\del \L},\o_{\L}, 
h_\L}_{\L}\left(m_{\L}\right)
=-\frac{1}{2 q}<m^*_\L(h_\L),R_{\L}m^*_\L(h_\L)>_{\L}
+\frac{1}{2}\sum_{x\in \L}(m^*_x(h_x))^2-\sum_{x\in \L}\eta_x(h_x)\cr
&-\frac{1}{q}<\tilde\eta_{\del(\L^c)}(q\tilde m),R_\L m^*_{\L}(h_\L)>_{\L}\cr
&\quad
-\frac{1}{2q}<\tilde\eta_{\del(\L^c)}(q\tilde m), 
R_\L\left(\tilde\eta_{\del(\L^c)}(q\tilde m)
\right)>_{\L}
+\frac{q}{2}\sum_{{x\in \L; y\in \del \L}
\atop{d(x,y)=1}}\tilde m_y^2\text{with}
\cr
&R_{\L}=\left(q^{-1}-\D_{\L} \right)^{-1}
}
\tag{3.1}
$$ 
with $\tilde\eta_{\del(\L^c)}(\tilde m):=
\del_{\L,\del\L}\tilde m_{\del\L}$ denoting the field
created by the boundary condition. 
We also note that the continuous-spin minimizer 
is given by (2.6).


We will deduce a contour representation for the 
$\nu$-measures. Let us give the commonly used 
notion of `contour', adapted to this model: 

{\bf Definitions: }{\it A {\bf contour} $\G$ in the volume $\L$ 
is a pair composed 
of a {\bf support} $\un\G\sb \L$ and a `height configuration' $h_{\L}\in {\Z}^{\L}$,
such that the extended 
configuration $(h_{\L},0_{\Z^d\ba \L})$ is 
constant on connected components of $\Z^d\ba \un\G$.
A {\bf contour model representation} for a probability measure 
$\nu$ on the space ${\Z}^{\L}$ of integer height-
configurations in $\L$ is a probability measure $Q$ 
on the space of contours in $\L$ whose height-marginal
reproduces $\nu$, i.e.
$\nu(\{h_{\L}\})= \sum_{{\G:}\atop {h_{\L}\left(\G\right)=h_{\L}}}
Q\left(\{\G\}\right)$.  
The {\bf connected components} of 
a contour $\G$ are the contours $\g_i$ whose 
supports are the connected components ${\un \g}_i$ of $\un \G$
and whose sign is determined by 
the requirement that it be the same as that of $\G$ on $\ov{{\un\g}_i}$.
}

The result of this chapter is

\proposition{1}{\it  
Suppose that $q$ is sufficiently small,
$q(m^*)^2$ sufficiently large and 
$\d_{d}\leq \frac{1}{4}$.
Then there  is a $h_{\L}$-independent constant 
$K_{\L}\left(\o_{\L}\right)$ s.t. we have  the representation 
$$
\eqalign{
&e^{-\inf_{m_{\L}\in \R^{\L}} H^{\tilde m_{\del \L}=0,\o_{\L}, 
h_\L}_{\L}\left(m_{\L}\right)}
=K_{\L}\left(\o_{\L}\right) \times e^{-<S(\o),V(h_{\L})>}
\sum_{{\G}\atop {h_{\L}\left(\G\right)=h_{\L}}}\r_0(\G; \o_{\un\G})
}
\tag{3.2}
$$
for any $h_{\L}\in \Z^{\L}$.
The quantities in the above representation are as follows:

\noindent{\bf (i) Small fields:} $S_{C}$ 
is a random variable for each $h\in \Z$ and $C\sb \Z^d$ connected
and we have used the notation  
$$
\eqalign{
&<S(\o), V(h_\L)>
:=\sum_{h\in \Z}
\sum_{C\sb \L\cap V_h}S_C(h)
\text{where} V_h(h_{\L}):=\{x\in \L, h_x=h\}         \cr
}
\tag{3.3}
$$
The $S_{C}(h)$ are functions of the random 
centerings $d_C(h)=(d_{x}(h))_{x\in C}$ for $|C|\geq 2$.
Up to a boundary term (see below), the single-site part
$\b S_x(h)$ is the `random depth' $\eta_x(h)$.

We have the smallness properties, 
for all realizations of the disorder,
$$
\eqalign{
&|S_{C}(h)|\leq \Const \d_d^2 
e^{-\const \a|C|}, \text{ for } |C|\geq 3
\text{ with } \a=\frac{1}{2}\log\left[1+1/(2dq)\right]\sim 
\frac{1}{2}\log\frac{1}{q}\cr
&|S_{C}(h)|\leq \Const \left(\d_d^2 +\d_{\eta}\right) 
\text{ for } |C|\leq 2
\cr
}
\tag{3.4}
$$
with $\const, \Const$ being of the order unity, depending
only on the dimension.

\noindent{\bf (ii) Contour-Activities:} 
The activity $\r_{0}(\G;\o_{\un\G})$  is non-negative.
It factorizes over the connected components of $\G$, i.e. we have
$\r_{0}(\G;\o_{\un \G})=
\prod_{\g_i\hbox{ conn cp. of }\G}
\r_{0}(\g_i;\o_{\g_i})$. 


\noindent
For $\un\G$ not touching the boundary (i.e. 
$\del_{\del \L}\un\G=\em$) the value of 
$\r_{0}(\G;\o_{\un\G})$ is independent of $\L$.
We then have the `infinite volume properties' 
of invariance under joint lattice shifts
of spins and random fields, as well as under joint
shift in the height-direction.

\noindent {\bf Peierls-type bounds:}
There exist positive constants $\tilde \b,\b$ s.t. 
we have the upper bounds
$$
\eqalign{
&0\leq \r_{0}(\G;\o_{\un \G})
\leq
e^{-\b E_s(\G)-\tilde \b |\un \G|}
\cr
}
\tag{3.5}
$$
with the `nearest-neighbor contour energy'
$$
\eqalign{
&E_s(\G):=
\sum_{{\{x,y\}\sb \ov{\un\G}}\atop{|x-y|=1}}|\tilde h_x-\tilde  h_y|
\text{ for a contour }\G=\left(\un \G,(\tilde h_x)_{x\in \L}\right)\cr
}
\tag{3.6}
$$

In the above estimates the `Peierls-constants' can be chosen like
$$
\eqalign{
&\b= q (m^*)^2 \frac{(1-2\d_d)^2}{12 [(1+2dq)^2-q^2]},\quad
\tilde \b=\Const \times \min\left\{
\log\frac{1}{q}, q{m^*}^2 \left(\frac{\log \frac{1}{q}}{\log m^*}
\right)^d\right\}\,\,\,\left( \leq \a,\b\right)
\cr
}
\tag{3.7}
$$
\noindent{\bf Probabilistic bounds for the small field:}
For $|C|=1,2$ we have $\E S_{C}=0$ and 
$$
\eqalign{
&\P\left[
\left|S_C(h)\right|\geq t
\right]\leq e^{-\frac{t^2}{2\s^2}}\text{with}
\s^2=\const (\s_{\eta}^2+\s_{d}^2)\cr
}
\tag{3.8}
$$
}

\bigskip\bigskip

\proof To produce a sum of the type (3.2) 
we need to decompose the terms in the Hamiltonian
in such a way as  to exhibit a low-temperature
part, a non-local field part and high-temperature
parts that can be expanded. 
We will produce `support of contours' from all these various
sources. As we will see there we will have to introduce 
some type of support
that occurs only for unbounded spin models with
interactions having no finite range.

Remember that we put zero boundary conditions.
First of all it is now convenient to rewrite the integer-height
Hamiltonian in the following form that makes 
explicit that it has purely ferromagnetic couplings:
$$
\eqalign{
&\inf_{m_{\L}\in \R^{\L}} H^{\tilde m_{\del \L}=0,\o_{\L}, 
h_\L}_{\L}\left(m_{\L}\right)\cr
&=\sum_{\{x,y\}\sb \ov \L} J_{\L;x,y}\left(\hat h_{x}-\hat h_{y}-\left[
\hat d_x(h_{x})-\hat d_y(h_{y})\right]\right)^2
-\sum_{x\in \L}\eta_x(h_x)+K_1\left(\o_{\L}\right)\text{with}\cr
&\hat h_x=h_x 1_{x\in \L},\quad\hat d_x(h)=d_x(h) 1_{x\in \L}
}
\tag{3.9}
$$
where $K_1(\o_{\L})$ is a constant that is independent 
of the height-configuration $h_{\L}$.
The $J_{\L}$'s are positive and their nearest neighbor 
parts satisfy the lower bound for nearest neighbors 
and and upper bound for the decay of the form 
$$
\eqalign{
&\min_{\{x,y\}\sb \ov{\L}, |x-y|=1}J_{\L; x,y}\geq 
q{m^*}^2 \left[4\left((1+2d q)^2 -q^2 \right)\right]^{-1}\cr
&J_{\L; x,y}\leq 
\frac{{m^*}^2(1+2dq)}{4} \left((2d q)^{-1}+1\right)^{-|x-y|}
}
\tag{3.10}
$$
where $|x-y|$ is the $1$-norm. 
Assuming this upper bound we have that 
$$
\eqalign{
&J_{\L; x,y}\leq e^{-\a|x-y|}\text{for }|x-y|\geq r(m^*,q)\text{where}\cr
&\a=\frac{1}{2}\log\left[1+1/(2dq)\right],
\quad r(m^*,q):=\left[2\frac{\log({m^*}^2(1+2dq)/4)}
{\log\left(1+1/(2dq)\right) }
\right]+1
}
\tag{3.11}
$$
where square brackets means integer part in the definition 
of the integer range $r=r(m^*,q)$.

Indeed, these couplings $J_{\L;x,y}$ can be conveniently read off 
from the following rewriting of the quadratic form
appearing in the minimum of the continuous-spin Hamiltonian
using the random walk representation of the resolvent: 
We decompose $R_{\L; x,y}=\sum_{C\sb \L}
\RR\left(x\rightarrow y\,; C\right)$ 
with
$\RR\left(x\rightarrow y\,; C\right)=
\sum_{\g}\left(q^{-1}+2d  \right)^{-(|\g|+1)}$
where the sum is over all nearest neighbor paths $\g$ on 
the lattice from $x$ to $y$ that visit precisely 
the connected set $C$ (see e.g. [K4]A.11 ff.) We have then 
$$
\eqalign{
&-<m_\L,R_{\L}m_\L>_{\L}
+q\sum_{x\in \L}m_x^2\cr
&=\sum_{x,y\in \L}\sum_{C\sb \L}\RR\left(x\,\rightarrow y\,; C\right) \frac{1}{2}\left[
\left(m_x-m_y \right)^2-m_x^2 - m_y^2
\right]
+\sum_{x\in \L}\sum_{C\sb \Z^d}\sum_{y\in \Z^d}\RR\left(x\,\rightarrow y\,; C\right)
m_x^2\cr
&=\frac{1}{2}\sum_{x,y\in \L}R_{\L;x,y}\left(m_x-m_y \right)^2
+\sum_{x\in \L}\left[\sum_{{C:C\cap (\Z^d\ba\L)\neq \em}\atop{y:y\in C}}
\RR\left(x\,\rightarrow y\,; C\right)\right]m_x^2
\cr
}
\tag{3.12}
$$
Indeed, this gives immediately the closely related form 
$$
\eqalign{
&\inf_{m_{\L}\in \R^{\L}} H^{\tilde m_{\del \L}=0,\o_{\L}, 
h_\L}_{\L}\left(m_{\L}\right)\cr
&=\sum_{\{x,y\}\sb \L} J_{\L;x,y}\left(h_{x}-h_{y}-\left[d_x(h_{x})-d_y(h_{y})\right]\right)^2
-\sum_{x\in \L}\eta_x(h_x)
+\sum_{x\in \L} K_{\L;x}\left(h_{x}-d_x(h_{x})\right)^2
+K_1(\o_{\L})
}
\tag{3.13}
$$
with 
$$
\eqalign{
&J_{\L;x,y}=\frac{ {m^*}^2}{4q}\sum_{C: x,y\in C\sb \L}
\RR\left(x\,\rightarrow y\,; C\right),
\quad
K_{\L;x}=\frac{{m^*}^2}{2q}\sum_{{C:C\cap (\Z^d\ba\L)\neq \em}\atop{y:y\in C}}
\RR\left(x\,\rightarrow y\,; C\right)\cr
}
\tag{3.14}
$$
To get the form (3.9) we can of course 
look at the boundary term as a coupling 
to a boundary condition $\tilde h_x\equiv 0$ for $x\in \del \L$.
Note that $K_{\L;x}$ falls off exponentially as a function 
of the distance from $x$ to $\del \L$.
Now, for every site $x\in \L$ we pick a site $y(x)\in \del \L$
that has minimal distance to $x$ (with some arbitrary deterministic
prescription to make this choice unique.)
Then we extend the definition
of $J$ to all pairs in $\Z^d\times \Z^d$
by   $J_{\L;x,y}:= K_{\L;x}$ for $y=y(x)$, 
$J_{\L;x,y}:=0$ for $y \in \left(\ov{\L} \right)^c$ or
$x,y \in \left(\L \right)^c$.
The lower bounds in (3.10) follow from the fact that
$\RR\left(x\rightarrow x+e\,; C\right)=
\left[(q^{-1}+2d)^2-1 \right]^{-1}$ for nearest neighbors $x,x+e$.
The upper bound follows from the fact that 
$$
\eqalign{
&\sum_{y\in \Z^d}\RR\left(x\,\rightarrow y\,; C\right)
\leq q\left(\frac{2d}{q^{-1}+2d}\right)^{|C|-1}
}
\tag{3.15}
$$
(see Appendix of [K4]).
Now, for a given configuration $h_{\L}$ there 
are pair interaction terms 
$J_{\L;x,y}\left(\hat h_{x}-\hat h_{y}
-\left[d_x(h_{x})-d_y(h_{y})\right]\right)^2$
that are {\it big} (they will make up the essential contributions
to the `low-temperature contours'), {\it small} (they will be expanded
and make up
high-temperature contours) and {\it intermediate}
(they cannot be expanded and will be adjoined
to the low-temperature contributions). 
The difficulty about these intermediate contributions
is that we need to find conditions
that ensure that their existence implies the existence of low-temperature 
contours nearby that will actually dominate them.  

Below we will introduce a set of pairs 
of `big and intermediate interactions',
$\EE_{\ov{\L}}(\hat h)\sb \ov{\L}\times \ov{\L}$.
In particular 
the pairs that make up the low-temperature contributions
will be contained in this set of `dangerous edges'.  
For fixed height configuration we decompose the exponential of (3.1):
$$
\eqalign{
&e^{-\sum_{\{x,y\}\sb \ov \L} J_{\L;x,y}\left(\hat h_{x}-\hat h_{y}-\left[
\hat d_x(h_{x})-\hat d_y(h_{y})\right]\right)^2
+\sum_{x\in \L}\eta_x(h_x)
}\cr
&=e^{-\sum_{\{x,y\}\in \EE_{\ov{\L}}(\hat h)} J_{\L;x,y}\left(\hat h_{x}-\hat h_{y}-\left[
\hat d_x(h_{x})-\hat d_y(h_{y})\right]\right)^2\,\,1_{\hat h_x\neq \hat h_y}}\cr
&\qquad\times e^{-\sum_{\{x,y\}\not\in \EE_{\ov{\L}}(\hat h)}J_{\L;x,y}\left(\hat h_{x}-\hat h_{y}-\left[
\hat d_x(h_{x})-\hat d_y(h_{y})\right]\right)^2\,\,1_{\hat h_x\neq \hat h_y}}\cr
&\quad\times e^{\sum_{h\in \Z}\left[-\left(
\sum_{\{x,y\}\sb \L\cap V_h} +
\sum_{{x\in\L\cap V_h}\atop{y\in \del \L}}
\right)
J_{\L;x,y}\left(\hat d_x(h)-\hat d_y(h)\right)^2
+\sum_{x\in V_h}\eta_x(h)
\right]}
}
\tag{3.16}
$$
The rest of the proof 
is a careful treatment the three exponentials on the r.h.s.
\bigskip

{\bf(i):} First exponential in (3.16): 
Low temperature-contours (nearest neighbor parts,
large fluctuation long range parts)

Given $h_{\L}$,
our aim is to define a support of a contour $\un{\G}^{\hbox{\srm LT}}(h_{\L})$,
s.t. the first exponential can be written as
a contour activity $\r^{\hbox{\srm LT}}
\left(\un{\G}^{\hbox{\srm LT}}(h_{\L}), h_{\L}\right)$
that satisfies a Peierls-type estimate
in terms of the n.n. surface energy {\it and}
the volume. 
We will have two contributions to the set of dangerous 
edges, $\EE_{\ov{\L}}(\hat h)
:=\EE^{(1)}_{\ov{\L}}(\hat h)\cup \EE^{(2)}_{\ov{\L}}(\hat h)$.
For the first, the {\bf short range part} we put
$$
\eqalign{
&\EE^{(1)}_{\ov{\L}}(\hat h):=
\{\{x,y\}\in \ov{\L}\times \ov{\L};
d(x,y)\leq r \hbox{ where }\hat h_x\neq \hat h_y 
\}
\cr
}
\tag{3.17}
$$
with the range  $r=r(q,m^*)\geq 1$  given above in (3.11). 
These interactions provide the `main-part' of the low-temperature
Peierls constant. Looking at these 
is in perfect analogy to what one would do
in the case of an Ising model. 
The corresponding ingredient of the support of the LT-Peierls contours
will be the connected components of the corresponding vertex set, i.e. 
$$
\eqalign{
&\un{\G}^{(1)}_{\ov{\L}}(\hat h):=
\{x\in \ov{\L};
\exists y\in \ov{\L}
\hbox{ s.t. }d(x,y)\leq r
\hbox{ where }\hat h_x\neq \hat h_y 
\}
\cr
}
\tag{3.18}
$$
Assuming that $\d\leq \frac{1}{4}$ it is readily seen that we have 
an energetic suppression in terms of the n.n. surface energy:
$$
\eqalign{
&\sum_{\{x,y\}\in\EE^{(1)}_{\ov{\L}}(\hat h)
} J_{\L;x,y}\left(\hat h_{x}-\hat h_{y}-\left[
\hat d_x(h_{x})-\hat d_y(h_{y})\right]\right)^2
\geq \t_{\hbox{n.n.}}\sum_{<x,y>\in \ov{\L}}|\hat h_x-\hat h_y|\text{with}\cr
&\qquad\t_{\hbox{n.n.}}:=(1-2\d_d)^2
\min_{\{x,y\}\sb \ov{\L}, |x-y|=1}J_{\L; x,y}
}
\tag{3.19}
$$


Now we come to the second LT-part of the 
support of the contour, the {\bf large fluctuation long range part}.
It is unavoidable since  the height variables 
can in principle have unbounded fluctuations: Indeed, if the difference
between heights at far away sites is extremely large,
it becomes impossible to treat the corresponding
terms in the Hamiltonian by an high-temperature expansion.
It is however intuitively clear that such an event 
should be very unlikely since it implies a large 
energy cost due to the short range parts.
To turn this into a contour 
representation we look at the following set of `dangerous'
bonds, 
$$
\eqalign{
&\EE_{\ov\L}^{(2)}(\hat h):=\{\{x,y\}\sb \ov{\L}\times  \ov{\L}; 
d(x,y)\geq r;\hbox{ where }|\hat h_x-\hat h_y|\geq e^{\frac{\a}{2}|x-y|}\}\cr
}
\tag{3.20}
$$
whose interactions 
can not be treated by a high-temperature expansion, 
with $\a$ being the estimate on the decay-rate of the $J$'s,
as given in (3.11).
The corresponding part of the LT-support will have to contain
the vertex sets of the
connected components of the corresponding graph.
Our aim is then to show the Peierls-type estimate
in terms of the n.n. interface energy {\it and }
the volume of the contour. 
The problem with the volume-estimate is that 
there is of coarse a gap between high-temperature 
and low-temperature expansions: if the interaction term 
$J_{\L;x,y}\left(\hat h_{x}-\hat h_{y}-\left[
\hat d_x(h_{x})-\hat d_y(h_{y})\right]\right)^2$
is not small enough for a high-temperature expansion, 
the term itself need not be large enough to provide a 
low-temperature Peierls constant that is big enough. 
So, what might happen is that the large fluctuation long range 
parts just glue together connected components 
of the $\un\G^{(1)}$-parts without contributing much
energy themselves. 
We will however show that in such a case the n.n. surface
energy will be at least as large as the resulting volume
of the total contour. 
Indeed, if the interaction term is not very small,
 each nearest neighbor path
from $x$ to $y$ contributes a nearest neighbor interface 
energy of $e^{\frac{\a}{2}|x-y|}$. Patching together those paths along with safety
cubes around them we will get the second part of the contour
with a useful Peierls constant. The details are as follows.

It is convenient to work in {\it all}
of ${\Z}^d$ and define sets that 
will be the essential part of the low-temperature contours
which are not necessarily subsets of $\L$. The final 
supports of contours will then be obtained as intersections.  
Given our extended configuration $\hat h$ 
it is convenient to extend the above definition writing
$\EE^{(2)}_{\Z^d}(\hat h):=\{\{x,y\}\sb {\Z}^d\times {\Z}^d; 
d(x,y)\geq r;\hbox{ where }|\hat h_x-\hat h_y|\geq e^{\a|x-y|}\}
$.
This is a finite set. 
To each pair $\{x,y\}\in \EE^{(2)}_{\Z^d}(\hat h)$ we associate
a cube $Q\left(\{x,y\}\right)\sb \Z^d$ among the cubes containing the points
$x$ and $y$ with the smallest side-length. (If the line between 
$x$ and $y$ doesn't happen to be a diagonal, we choose some arbitrary
deterministic tie-breaking procedure to make this choice unique.)
Then we put
$\un{\G}^{(2)}_{\Z^d}(\hat h)
:=\bigcup_{\{x,y\}\in\EE^{(2)}_{\Z^d}(\hat h)}Q\left(\{x,y\}\right)$.  
Finally we define the LT-support $\un{\G}^{\hbox{\srm LT}}(h)\sb \L$ 
of the contour by
$$
\eqalign{
&\un{\G}^{\hbox{\srm LT}}(h):=
\left(\un{\G}^{(1)}_{\ov{\L}}(\hat h)
\cup \un{\G}^{(2)}_{\Z^d}(\hat h)\right)\cap \L
\cr
}
\tag{3.21}
$$
Then the desired lower bound of the n.n. surface energy
in terms of the  volume of the long-range parts is 
given by the following

\lemma{3.1}{\it
Suppose that $\un\g$ is a connected component of 
the set $\un{\G}^{(2)}_{\Z^d}(\hat h)$. Then the nearest neighbor
surface energy satisfies a Peierls estimate of the form 
$$
\eqalign{
&\sum_{<x,y>\in \un\g}|\hat h_x-\hat h_y|\geq K\left|\un\g\right|
\text{with}K:=\inf_{L'\geq r}\left(\frac{e^{\frac{\a L'}{2}}}{(3 L')^d}\right)
\geq \left(\frac{e\a}{6d}\right)^d    
}
\tag{3.22}
$$
}

\proof
From the set of cubes $Q(\{x,y\})$
whose union  makes up $\un\g$ we will consider in the following 
only the maximal ones w.r.t. inclusion. Let us denote them by $Q_i$, 
$i=1,\dots, N$.  (That is we discard those that are contained 
in a strictly bigger one.)
We prove the Lemma by induction over the number $N$ of such cubes of a connected component. 
In the case of one cube, say $Q(x,y)$ of side-length $L$, we have that 
$$
\eqalign{
&\sum_{<i,j>\in Q(x,y)}|h_i-h_j|\geq |h_x-h_y|\cr
&\geq e^{\frac{\a}{2}|x-y|}
\geq e^{\frac{\a}{2} L} \geq L^d\inf_{L'\geq r}\left(\frac{e^{\frac{\a}{2} L'}}{{L'}^d} \right)
=3^d K^d L^d
}
\tag{3.23}
$$
where the first inequality is the triangle inequality,
the second the definition of the `dangerous bonds'.
This proves the single-cube case even with a constant $3^d K$.

Now, given a contour $\un\g=\bigcup_{i=1}^N Q_i$,
we define a smaller contour $\un\G'$ in the following way.
Pick one of the cubes $Q_i$ with the largest side-length, call this cube
$Q_{i_0}$ (with corresponding side-length $L_{i_0}$).
Define 
$$
\eqalign{
&\un \G':=\bigcup_{{i\in\{1,\dots,N\}; i\neq i_0}\atop
{Q_i\cap Q_{i_0}}=\em}Q_i
}
\tag{3.24}
$$
The resulting contour might be connected or not.
Since we took away the {\it biggest} cube and its `contact partners' 
we have $\left|\un\G' \right|\geq \left|\un\g \right|- 3^d L_{i_0}^d$.
For the nearest neighbor surface energy we have  
$$
\eqalign{
&\sum_{<x,y>\in \un\g}|h_x-h_y|\geq \sum_{<x,y>\in \un\G'}|h_x-h_y|
+\sum_{<x,y>\in Q_{i_0}}|h_x-h_y|
}
\tag{3.25}
$$
Using the induction hypothesis on the connected components of $\un\G'$
and the estimate on the single cube from above we get the desired 
$$
\eqalign{
&\sum_{<x,y>\in \un\g}|h_x-h_y|\geq 
K\left|\un\G'\right|+ 3^d K L_{i_0}^d\geq K\left|\un\g \right|
}
\tag{3.26}
$$
\endproof

\bigskip

Now we are done with the treatment of the first exponential
in (3.16): Indeed, the sum 
$\sum_{\{x,y\}\in \EE_{\ov{\L}}(\hat h)} J_{\L;x,y}\left(\hat h_{x}-\hat h_{y}-\left[
\hat d_x(h_{x})-\hat d_y(h_{y})\right]\right)^2\,\,1_{\hat h_x\neq \hat h_y}
$
decomposes over connected components ${\un\g}_{i}$ of 
$\un{\G}^{\hbox{\srm LT}}(h)$. 
Denote the corresponding pairs by $\EE_{\ov{\L};i}(\hat h)$ (i.e.
the pairs with vertices in $\un \g\cup \left(\ov{\un \g}\cap \del \L\right)$).
Denote by $\g_{i}$ the contour whose support is $\un \g_i$ and whose
height configuration is $\hat h_{\ov{\L}}$ on $\ov{{\un\g}_i}$.
We define the LT-activities by  
$$
\eqalign{
&\r^{\hbox{\srm LT}}(\g_i)
:=e^{-
\sum_{\{x,y\}\in \EE_{\ov{\L};i}(\hat h)} J_{\L;x,y}\left(\hat h_{x}-\hat h_{y}-\left[
\hat d_x(h_{x})-\hat d_y(h_{y})\right]\right)^2\,\,1_{\hat h_x\neq \hat h_y}}
\cr
}
\tag{3.27}
$$
The desired LT surface-energy/volume Peierls-type estimate is obvious: 
Denote ${\un \G}^{(2)}:=\g^{(i)}\cap \un{\G}^{(2)}_{\ov{\L}}(\hat h)$
the parts of the connected component that are due to the 
large-fluctuation-long-range part. Then we have 
$$
\eqalign{
&\sum_{\{x,y\}\in \EE_{\ov{\L};i}(\hat h)} J_{\L;x,y}\left(\hat h_{x}-\hat h_{y}-\left[
\hat d_x(h_{x})-\hat d_y(h_{y})\right]\right)^2\,\,1_{\hat h_x\neq \hat h_y}
\geq \t_{\srm n.n.} E_s(\g_i)\cr
&\geq \frac{\t_{\srm n.n.}}{3} E_s(\g_i)
+\frac{\t_{\srm n.n.}}{3}\frac{\left|{\un\g}_i \cap 
\un{\G}^{(1)}_{\ov{\L}}(\hat h)\right|}{(2r+1)^d}
+\frac{\t_{\srm n.n.}}{3} K\left|{\un\g}_i \cap 
\un{\G}^{(2)}_{\ov{\L}}(\hat h)\right|\cr
&\geq \b E_s(\g_i)
+\t_1 \left|{\un\g}_i\right|
}
\tag{3.28}
$$
\bigskip 
with $\b=\frac{\t_{\srm n.n.}}{3}$ and $\t_1=\frac{\t_{\srm n.n.}}{3}
\times \min\{(2r+1)^{-d},K\}$.
Let us look at the large $m^*$, small $q$-asymptotics with
$\t_{\srm n.n.}\sim q{m^*}^2/4$ sufficiently large,
$\a\sim\frac{1}{2}\log\frac{1}{q}$,
$r\sim 4 \frac{\log m^*}{\log\frac{1}{q}}$.
From this we get that $\b\sim  q{m^*}^2/12$,
$\t_1\sim  \Const q{m^*}^2 
\left(\frac{\log\frac{1}{q}}{\log m^*}\right)^d $.
(Note that the contributions
of the $K$-term will actually obtain a better Peierls
constant that hence won't be visible.) 

{\bf (ii):} Second exponential in (3.16): High-temperature expansion

Let us just write $\un\G:=\un{\G}^{\hbox{\srm LT}}(h)$.
We find it convenient to use a little rewriting of the exponent. 
Since we have exponential decay of the interaction, 
we can just `fill the space between the endpoints' to define contours
that obey Peierls estimates:
To each pair $\{x,y\}\in \left[\ov{\L}\times \ov{\L}\right]
\ba\EE_{\ov{\L}}(\hat h)$ 
of sites we associate a `one-dimensional' polymer
$g=g(x,y)\sb \ov{\L}$ that is the set of sites of one of the nearest-neighbor 
paths from $x$ to $y$ (with some prescription to make the choice 
of this path unique.)
Then we have for the number of sites that $|g|=|x-y|_1+1$.
We use this notation to denote the terms in the last sum by sets $g$ and put
$$
\eqalign{
&S_g(h_g):=1_{\{x,y\}\not\in \EE_{\ov{\L}}(\hat h)}
\times J_{\L;x,y}\left(\hat h_{x}-\hat h_{y}-\left[
\hat d_x(h_{x})-\hat d_y(h_{y})\right]\right)^2\,\,1_{\hat h_x\neq \hat h_y}
}
\tag{3.19}
$$
We note that, due to the decay of the resolvent (with $\a$), the uniform 
boundedness of $|d_x|\leq \d_d$ and the definition of the `dangerous bonds'
(with $\a/2$!) we have 
$0\leq S_g(h_g)\leq e^{-\const\a|g|}$.
Note that there are only non-vanishing terms for $g\cap \un \G\neq \em$.
(Indeed, in the case that both $x$ and $y$ are not in $\un \G$, they must
lie in different connected components of the complement of $\un \G$.)
The exponential can now be treated by the subtraction of bounds trick:
We have 
$$
\eqalign{
&e^{-\sum_{g:g\cap \un \G\neq \em}S_g(h_g)}
=\prod_{\un\g\hbox{ \srm conn.cp. of }\un\G}e^{-\sum_{g:g\cap \un \g\neq \em}
e^{-{\const\a|g|}}}
\times e^{\sum_{g:g\cap \un \G\neq \em} 
\left(n(\un\G,g)e^{-{\const\a|g|}}- S_g(h_g)   \right)}
}
\tag{3.30}
$$
where $n(\un\G,g)$ counts the number of connected components
of $\un \G$ that are connected to $g$.
The term under the first product defines a non-negative quantity 
$r(\un \g)$ that is $h$-independent and satisfies
$1\geq r(\un \g)\geq e^{-|\un\g|e^{-\const' \a}}$ (for suff. large $\a$).
The last exponential can be polymer-expanded and written as a sum
$\sum_{G:G\cap \un \G\neq \em}\r^{\srm HT1}_{\un\G}(G,h_{G})$
with a nonnegative activity $0\leq 
\r_{\un\G}(G,h_{G})\leq e^{-\const'\a|G|}$
(which is one for empty $G$).


\bigskip
{\bf (iii):} Third exponential in (3.16): Non-local small fields

The last exponential would not be present in the corresponding 
model without randomness. It describes the random
modulations of the `vacuum-energy' caused by the $d$-variables
in the flat pieces  outside the LT-contours (where 
height-fluctuations are occuring).  
To get the decomposition 
into $S_{C}$'s we use  
the decomposition of the resolvent and define for all $C$'s
$$
\eqalign{
&\b \tilde S_{C}(h_C)
:=\frac{{m^*}^2}{4q}\sum_{{x,y\in C:}\atop{h_x=h_y}}
\RR\left(x\,\rightarrow y\,; C\right)
\left(d_{x}(h_x)-d_{y}(h_y) \right)^2
\cr
}
\tag{3.31}
$$
From the bound (3.15) we get 
$$
\eqalign{
&\left|\b \tilde S_{C}(h_C)\right|
\leq (2\d_d)^2 q(m^*)^2
\frac{2d}{1+2dq} e^{-\a(|C|-2)}\cr
}
\tag{3.32}
$$
This is always fine for $|C|\geq 3$; 
for $|C|=2$ we see that $\d_d$ really needs to be small enough
to get a useful bound. 
We put $S_{C}:= \tilde S_{C}$ for $|C|\geq 3$
and $S_{C}:= \tilde S_{C}- \E\tilde S_{C}$
for $|C|=2$. 
[We remark 
that we could relax the assumption
of smallness of $\d_d$ by the introduction
of large 
$d_{x}(h)=d_{x}(h)1_{|d_{x}(h)|\geq \d_1}
+d_{x}(h)1_{|d_{x}(h)|< \d_1}=:d_{x}^{\hbox{\srm l}}(h)
+d_{x}^{\hbox{\srm s}}(h)$.
Then we would have to introduce a control field $N_x(h)$
that would contain a contribution of the 
type $\Const \left|d_{x}(h)\right| 1_{|d_{x}(h)|\geq \d_1}$
as well as a contribution from the large $\eta_x(h)$'s. 
Since this is simple but would obscure 
the structure of the contour-model we don't
present the details here.]

For 
a flat height configuration on $C$ (i.e. if $C\sb V_h$ 
for an $h\in \Z$) we just write $S_{C}(h):=
S_{C}(h_C)$ 
the former defining the `small field' appearing in the 
final contour-model representation. 
Then we have for the bulk term in the third
exponential in (3.16)
$$
\eqalign{
&\sum_{h\in \Z}\sum_{\{x,y\}\sb \L\cap V_h}
J_{\L;x,y}\left(d_x(h)-d_y(h)\right)^2
=\b \sum_{h\in \Z}\sum_{C\sb \L\cap V_h}S_C(h)
+\b \sum_{{C\sb \L}\atop {h_C\not\equiv \const}}S_C(h_C)
+K'_{\L}\cr
}
\tag{3.33}
$$
with some obvious $h_{\L}$-independent constant $K'_{\L}$.
The first term is the desired small-field contribution.
The second term is attached to the LT-contours and can 
be expanded. 
Using subtraction-of-bounds as before, its expansion gives
$$
\eqalign{
&e^{-\b \sum_{{C\sb \L}\atop {h_C\not\equiv \const}}S_C(h_C)}
=\left[\prod_{\un\g\hbox{ \srm conn.cp. of }\un{\G}^{\hbox{\srm LT}}(h)}
r_2(\un \g)\right]
\sum_{G:G\cap\un{\G}^{\hbox{\srm LT}}(h)\neq \em}
\r^{\srm HT2}_{\un{\G}^{\hbox{\srm LT}}(h)}(G,h_{G})\cr
}
\tag{3.34}
$$
with $\const\a$-decay. 
As far as for the bulk terms, the fields $\eta_x(h)$ simply
make up the local contributions to the small field. 
Finally, we discuss the corrections due to boundary effects. 
We write the interaction with the boundary in the form 
$$
\eqalign{
&\sum_{h\in \Z}\sum_{{x\in\L\cap V_h}\atop{y\in \del \L}}
J_{\L;x,y}\left(\hat d_x(h)-\hat d_y(h)\right)^2
=-\sum_{x\in \L}\tilde\eta_x(h_x)
+\tilde K_{\L}\text{where}\cr
&\tilde\eta_x(h)
:=-K_{\L;x}\left({d_x(h)}^2-\E\left[{d_x(h)}^2 \right]
\right)
}
\tag{3.35}
$$
We note that 
$\left|\tilde\eta_x(h)\right|
\leq q{m^*}^2\frac{d \d_d^2}{1+2dq}
$.
The centered field $\tilde\eta_x(h)$ 
will just give a small volume dependent modification 
of the local small field that we finally define by
$$
\eqalign{
&S_x(h):=\eta_{x}^{\hbox{\srm s}}(h)+\tilde\eta_x(h)\cr
}
\tag{3.36}
$$
We note that from this definition the 
probabilistic bounds for $|C|=1$ are clear.
Also, the probabilistic bounds for $|C|=2$ 
are obvious.

\medskip
Finally putting together our results from that  (i)-(iii)
we end up with the representation 
$$
\eqalign{
&e^{-\sum_{\{x,y\}\sb \ov \L} J_{\L;x,y}\left(\hat h_{x}-\hat h_{y}-\left[
\hat d_x(h_{x})-\hat d_y(h_{y})\right]\right)^2
+\sum_{x\in \L}\eta_x(h_x)
}\cr
&=\prod_{\un\g\hbox{ \srm conn.cp. of }\un{\G}^{\hbox{\srm LT}}(h_{\L})}
\left(r(\un \g)r_2(\un \g)
\r^{\hbox{\srm LT}}(\un\g,h_{\un\g})
\right)
\sum_{{G:G\cap \un{\G}^{\hbox{\srm LT}}(h_{\L})\neq \em}\atop
{G_2:G_2\cap \un{\G}^{\hbox{\srm LT}}(h_{\L})\neq \em}
}\r^{\srm HT1}_{\un{\G}^{\hbox{\srm LT}}(h_{\L})}(G,h_{G})
\r^{\srm HT2}_{\un{\G}^{\hbox{\srm LT}}(h_{\L})}(G_2,h_{G_2})\cr
&\times e^{-<S,V(h_{\L})>}
}
\tag{3.37}
$$
Resumming over $G,G_2$'s  that have
the same 
$\un{\G}^{\hbox{\srm LT}}(h_{\L})\cup\un{\G}^{D}(h_{\L})
\cup G\cup G_2=:
\un\G$ we get the desired form.
This concludes the proof of the proposition. 
\endproof

\bigskip\bigskip

\chap{IV. Proof of Theorem 1}

The proof is a direct consequence
of the representation of Theorem 2,
the contour-representation for the discrete-height
model of Proposition 1,
and the results from the renormalization
group analysis for a discrete-height contour 
model from [BoK1], [K1].
It is crucial for this that the contour model constructed
in Chapter III and 
given in Proposition 1 is `renormalizable' 
with the procedure described in detail in [BoK1] in 
`Chapter 4. The Gibbs State at Finite Temperature'.
Indeed, it satisfies the inductive assumption of 
a contour model given in [BoK1] 4.1, p. 457, 
for the trivial choice of empty bad regions
and vanishing control-field $N$.
We must however have for this that the uniform bounds
$\d_{\eta}$ and $\d_{d}$ are sufficiently small; 
otherwise we would have had to introduce
bad regions and large fields, which is however
mainly a notational inconvenience. 
(There is further a completely 
trivial difference in that we have 
exponential decay for the small fields 
only for $|C|\geq 3$; 
we could of course trivially cast the present
contour representation into the one from [BoK1]
by splitting the field $S_{C}$ for $C=\{x,x+e\}$
into new local small fields at sites $x$ and $x+e$ 
and producing a stochastic dependence up to distance $2$.)
The result of [BoK1] then gives that, for 
sufficiently large $\tilde \b\,\,(\leq \a, \b)$,
for sufficiently small $\s^2_{\hbox{\srm eff.}}$
there exists a non-random subsequence of cubes $\L$
s.t. the measures $\nu_{\L}^{\o}$ (obtained from 
the zero boundary continuous Gibbs-measures)
converge weakly to an infinite-volume 
Gibbs-measure $\nu^{\o}$, for a.e. $\o$ [see [BoK1], p.417,
Theorem 1].
To conclude that convergence of the $\mu$-measures takes 
place also on local observables $f$ that 
are only polynomially bounded, $|f(m_V)|\leq \Const(1+|m_V|)^p$,
we would like to use the addition to Theorem 2 (2.7). Although
its assumption on the convergence of the 
$\nu$-measures on exponentially 
bounded observables is a very natural one 
that is believed to hold, it is unfortunately
not a straightforward consequence of the RG-analysis.
Along the lines of Chapter II, 
the reader will however have no difficulty 
to prove the analogous extension for polynomially 
bounded observables under the condition
that $\sup_{\L,y} \E\nu_{\L}|h_y|^p<\infty$, for all exponents
$p$. This assumption is in fact true; we even have (4.2), see below.

Let us use the short notation 
$\NN[h_{\Z^d}]:=
\NN\left[\left(1-q\D_{\Z^d} \right)^{-1}m^*_{\Z^d}
\left(h_{\Z^d} \right)
;\left(1-q\D_{\Z^d}\right)^{-1}
\right]$. Then we have for the second moment
$$
\eqalign{
&\E\mu^{\o}
\left(m_{x_0}^2\right)
=\E\nu^{\o}
\left\{
\NN[h_{\Z^d}]
\left(m_{x_0}^2\right)
-\NN[h_{\Z^d}]\left(m_{x_0}\right)^2\right\}
+\E\nu^{\o}\NN[h_{\Z^d}]\left(m_{x_0}\right)^2\cr
&= \left(1-q\D_{\Z^d}\right)^{-1}_{0,0}
+\E\nu^{\o}\left(\sum_{y}\left(1-q\D_{\Z^d} \right)^{-1}_{x_0,y}
m^*_{y}(h_y)\right)^2
}
\tag{4.1}
$$
With the Schwartz inequality we bound the last expectation
by $(m^*)^2\sup_{y}
\E\nu^{\o}\left[\left(|h_y|+\d_{d}\right)^2\right]$.
To estimate this last expectation, we
utilize a corollary of the RG-analysis for 
the discrete height model of [BoK1] (whose
complete proof can be found in [K1], see Theorem 1.2 therein) 
saying that, for any $q\in \Z$, we have
$$
\eqalign{
&\E\nu^{\o}\left[|h_{y}|^q \right]\leq K(q)\left( e^{-\frac{1}
{\s^{\k}_{\hbox{\srm eff.}}}}
+e^{-\Const \tilde \b}
\right)
}
\tag{4.2}
$$
for any $y$,
with $q$-dependent $K(q)$ and some positive exponent $\k$.
This immediately proves (1.4).\endproof

To get more estimates on the continuous infinite volume
measures in terms of the discrete one might now utilize 
$\mu^{\o}[B]\leq \nu^{\o}[A^c]
+\sup_{h_{\Z^d}\in A}\NN\left[h_{\Z^d}\right][B]$
for any good choice 
of an event $A$ in integer height space.



\ftn
\font\bf=cmbx8

\baselineskip=10pt
\parskip=4pt
\rightskip=0.5truecm
\bigskip\bigskip
\chap{References}
\medskip


\item{[AW]} M.Aizenman, J.Wehr, Rounding Effects of Quenched Randomness
on First-Order Phase Transitions, Comm. Math.Phys {\bf  130},
489-528 (1990)

\item{[Ba1]} Balaban, Tadeusz,
A low temperature expansion for classical N-vector models. 
I. A renormalization group
flow, Comm.Math.Phys. {\bf 167} no. 1, 103--154 (1995)

\item{[Ba2]} Balaban, Tadeusz,
A low temperature expansion for classical N-vector models. 
II. Renormalization group
equations, Comm.Math.Phys. {\bf 182} no. 3, 675--721 (1996)




\item{[BCJ]} C.Borgs, J.T.Chayes, J.Fr\"olich,
Dobrushin States for Classical Spin Systems with 
Complex Interactions,
J.Stat.Phys. {\bf 89} no.5/6, 895-928 (1997)

\item{[BI]} E.Bolthausen, D.Ioffe, 
Harmonic crystal on the wall: a microscopic approach,
Comm.Math.Phys. {\bf 187} no.3, 523-566 (1997)

\item{[BK1]} J.Bricmont, A.Kupiainen, 
Phase transition in the 3d random field Ising model,
Comm.
Math.Phys. {\bf 142}, 539-572 (1988)


\item{[BKL]} J.Bricmont, A.Kupiainen, R. Lefevere,
       Renormalization Group Pathologies and the 
Definition of Gibbs States,
Comm. Math.Phys. {\bf 194} 2, 359-388 (1998)
    
\item{[BoG]}
A.Bovier, V.Gayrard, Hopfield models as generalized
random mean field models, in: 
Mathematical aspects of spin glasses and neural networks,
Progr. Probab. 41, 3-89, Birkh\"auser Boston, Boston, MA, 1998. 

\item{[BoK1]} A.Bovier, C.K\"ulske, A rigorous
renormalization group method for interfaces in random
media, Rev.Math.Phys. {\bf 6}, no.3, 413-496 (1994)

\item{[BoK2]} A.Bovier, C.K\"ulske, 
There are no nice interfaces in $2+1$ dimensional 
SOS-models in random media, J.Stat.Phys. {\bf 83}, 751-759
(1996)

\item{[B]} D.Brydges, A short course on cluster expansions, 
in `Critical phenomena, random systems, gauge theories' 
(Les Houches 1984) (K.Osterwalder, R.Stora, eds.),
North Holland, Amsterdam (1986)



\item{[Do1]} R.L.Dobrushin, 
Gibbs states describing a coexistence of phases 
for the three-dimensional Ising model,
Th.Prob. and its Appl. {\bf 17}, 582-600 (1972)


\item{[DoZ]} R.L.Dobrushin, M.Zahradnik, Phase Diagrams of Continuum Spin 
Systems, Math.Problems of Stat.Phys. and Dynamics, ed.
R.L.Dobrushin, Reidel pp.1-123 (1986)




\item{[EFS]} A.C.D.van Enter, R. Fern\'andez, A.Sokal,
Regularity properties and pathologies of position-space
renormalization-group transformations: Scope
and limitations of Gibbsian theory. J.Stat.Phys.
{\bf 72}, 879-1167 (1993)

\item{[F]} R. Fernandez, Measures for lattice systems,
preprint, to appear in: Proceedings of Statphys 20, Paris (1998),
available as preprint 98-567 at http://www.ma.utexas.edu/mp\_arc

\item{[FFS]} R.Fernandez, J.Fr\"ohlich, A.Sokal,
Random Walks, Critical Phenomena, and Triviality
in Quantum Field Theory, 
Springer, Berlin, Heidelberg, New York (1992) 

\item{[Geo]} H.O. Georgii, Gibbs measures and phase transitions, Studies
in mathematics, vol. 9 (de Gruyter, Berlin, New York, 1988)


\item{[K1]} C.K\"ulske, Ph.D. Thesis, Ruhr-Universit\"at Bochum (1993)

\item{[K2]} 
C.K\"ulske, A random energy model for size dependence:
recurrence vs. transience, Prob.Theor.Rel.Fields {\bf 111}, 
57-100 (1998)

\item{[K3]} 
C.K\"ulske, Limiting behavior of random Gibbs measures: metastates 
in some disordered mean field
models, in: 
Mathematical aspects of spin glasses and neural networks,  
Progr. Probab. {\bf 41}, 151-160,
eds. A.Bovier, P.Picco, Birkh\"auser Boston,
Boston (1998)

\item{[K4]} 
C.K\"ulske, The continuous spin 
random field model: Ferromagnetic ordering in $d\geq 3$,
to be published in Rev.Math.Phys, available at 
http://www.ma.utexas.edu/mp\_arc/, prepint 98-175 (1998)


abstract polymer models, Comm.Math.Phys. {\bf 103},
491-498 (1986)

\item{[Z1]} M.Zahradn\' \i k, On the structure of low temperature
phases in three dimensional spin models with random impurities:
A general Pirogov-Sinai approach, Mathematical Physics
of Disordered Systems, Proc. of a workshop held at the CIRM,
ed. A.Bovier and F.Koukiou, IAAS-Report No.4, Berlin (1992)

\item{[Z2]} M.Zahradn\' \i k, Contour Methods and Pirogov
Sinai Theory for Continous Spin Models,
preprint Prague (1998),
to appear in the AMS volume dedicated to R.L.Dobrushin

\end